**Price, Volatility and the Second-Order Economic Theory**

Victor Olkhov

TVEL, Moscow, Russia

victor.olkhov@gmail.com

ORCID iD 0000-0003-0944-5113

## Abstract

We introduce the price probability measure *η(p;t)* that defines the mean price *p(1;t)*, mean square price *p(2;t),* price volatility $\sigma_p^2(t)$ and all price *n-th* statistical moments *p(n;t)* as ratio of sums of *n-th* degree values *C(n;t)* and volumes *U(n;t)* of market trades aggregated during certain time interval *Δ*. The definition of the mean price *p(1;t)* coincides with definition of the volume weighted average price (VWAP) introduced at least 30 years ago. We show that price volatility $\sigma_p^2(t)$ forecasting requires modeling evolution of the sums of second-degree values *C(2;t)* and volumes *U(2;t)*. We call this model as second-order economic theory. We use numerical continuous risk ratings as ground for risk assessment of economic agents and distribute agents by risk ratings as coordinates. We introduce continuous economic media approximation of squares of values and volumes of agents trades and their flows aggregated during time interval *Δ*. We take into account expectations that govern agents trades and introduce aggregated expectations alike to aggregated trades. We derive equations for continuous economic media approximation on the second-degree trades. In the linear approximation we derive mean square price *p(2;t)* and volatility $\sigma_p^2(t)$ disturbances as functions of the first and second-degree trades disturbances. Description of each next *n-th* price statistical moment *p(n;t)* with respect to the unit price measure *η(p;t)* depends on sums of *n-th* degree values *C(n;t)* and volumes *U(n;t)* of market trades and hence requires development of the corresponding *n-th* order economic theory.

Keywords: volatility, economic theory, market transactions, expectations, price probability

JEL : C1, D4, E3, E4, G1, G2

---

This research did not receive any assistance, specific grant or financial support from TVEL or funding agencies in the public, commercial, or not-for-profit sectors. We appreciate proposals to fund our studies.

## 1. Introduction

"In everyday language, volatility refers to the fluctuations observed in some phenomenon over time" (Andersen et.al., 2005). Price and returns volatility and their impact on economy are the core issues of financial economics for decades: Hall and Hitch (1939), Fama (1965), Stigler and Kindahl (1970), Pearce (1983), Tauchen and Pitts (1983), Schwert, (1988), Mankiw, Romer and Shapiro (1991), Campbell et.al. (1993), Heston (1993), Brock and LeBaron (1995), Bernanke and Gertler (1999), Andersen et.al. (2001), Engle and Patton (2001), Poon and Granger (2003), Andersen et.al. (2005), Daly (2008), Christiansen, Schmeling and Schrimpf (2012), Padungsaksawasdi and Daigler (2018), Bogousslavsky and Collin-Dufresne (2019). We refer a small part of infinite number of studies to outline importance and complexity of the price-volatility problems.

Volatility is defined as standard deviation of certain probability distribution. Each new volatility model considers some specific properties of the proposed probability distribution and describes corresponding properties of the standard deviations. The choice and financial justification of the probability distribution that models the market random price dynamics is the core issue of the most studies. This approach separates studies of the price uncertainties from studies of the financial markets itself. Nevertheless a lot of papers investigate correlations and mutual dependence between volatility, volume and number of market trades (Tauchen and Pitts 1983; Campbell et.al. 1993; Ito and Lin 1993; Avramov et.al. 2006; Ciner and Sackley 2007; Miloudi et.al. 2016; Takaishi and Chen 2017).

It is important to underline that any model that describes random financial market properties models economic and financial variables aggregated during certain time interval $\Delta$. This interval $\Delta$ can be equal minute, hour, day, month and etc. Any description of price volatility uses financial data aggregated during some time interval $\Delta$. The choice of interval $\Delta$ and the choice of the aggregation method of the economic and financial data, price time-series, assets, investment, credits, profits and etc., plays core role for adequacy of the macroeconomic modeling. The duration of interval $\Delta$ determines important properties of the economic model – its accuracy, sustainability, scales of possible fluctuations and etc. Choice of time interval $\Delta$ and choice of data aggregation method are the most important initial steps of economic forecasting.

Let's note that modern macroeconomic and financial theories describe evolution of first-degree macro variables, transactions and expectations aggregated during certain interval $\Delta$. For example, macro investment, credits, assets are determined as sum of investment, credits and assets of economic agents (without doubling) during time interval $\Delta$. The sum of trades during interval $\Delta$ define macro trades, sum of consumption, taxes, debts define total consumption, total taxes, total debts in the economy aggregated during interval $\Delta$. All these macro variables are defined as sum of first-degree variables. We call description of macro first-degree variables and trades as the first-order economic theory. Almost all macroeconomic and financial variables are the first-degree variables.

Price and returns volatility are almost the only exception. Indeed, price volatility describes difference between mean square price and square of mean price averaged during certain interval $\Delta$. We outline that the price averaging procedures that determine the mean price and the mean square price – are the origin of the important economic problems. Actually financial markets and researchers use different averaging methods to define mean price.

The most common notion to measure price $p$ probability and mean price of market trades during time interval $\Delta$ based on the assumption that all trades have equal probability. Thus for the case with $N$ trades performed during averaging interval $\Delta$ probability of single trade is taken as $1/N$. If there are $n_p$ trades with price $p$ then probability $P(p)$ of the price $p$ and mean price $E(p)$ equals

$$P(p) = \frac{n_p}{N} \quad ; \quad E(p) = \sum p_i P(p_i) = \frac{1}{N} \sum p_i n_i \quad ; \quad N = \sum n_i \tag{1.1}$$

Here $n_i$ define number of trades with price $p_i$. Relations (1.1) are so common and simple that almost no additional considerations support assumption (1.1). However during last decades a different definition of the mean price was proposed. The volume weighted average price (VWAP) $p(1;t)$ was introduced and studied by (Berkowitz et.al 1988; Buryak and Guo, 2014; Guéant and Royer, 2014; Busseti and Boyd, 2015; Padungsaksawasdi and Daigler, 2018, CME Group, 2020).

$$C(1;t) = \sum_{i=1}^{N(t)} C(t_i) \quad ; \quad U(1;t) = \sum_{i=1}^{N(t)} U(t_i) \quad ; \quad C(t_i) = p(t_i)U(t_i) \tag{1.2}$$

$$p(1;t) = \frac{1}{U(1;t)} \sum_{i=1}^{N(t)} p(t_i)U(t_i) = \frac{C(1;t)}{U(1;t)} \tag{1.3}$$

$$\Delta(t) = \left[t - \frac{\Delta}{2}, t + \frac{\Delta}{2}\right] \quad ; \quad t_i \in \Delta(t)\,, \; i = 1, \dots N(t) \tag{1.4}$$

$$\mu(t_i) = \frac{U(t_i)}{U(1;t)} \quad ; \quad p(1;t) = \sum_{i=1}^{N(t)} p(t_i)\mu(t_i) \quad ; \quad \sum_{i=1}^{N(t)} \mu(t_i) = 1 \tag{1.5}$$

Here $C(t_i)$ denote the value and $U(t_i)$ the volume of $N(t)$ trades with price $p(t_i)$ performed at moment $t_i$ during averaging interval $\Delta(t)$ (1.4). VWAP $p(1;t)$ defines the mean price as the ratio of sum of the values $C(t_i)$ to sum of the volumes $U(t_i)$ of all $N(t)$ market trades during interval $\Delta$. One may say that $p(1;t)$ VWAP is averaged by unit measure $\mu(t_i)$ (1.5).

The collision between two definitions of the mean price hides difference between two approaches to description of irregular price evolution. The common approach regards price as random function and (1.1) use frequencies $n_p/N$ of trades with price $p$ to describe price probability, mean price, volatility etc. But investors don't like gambling and prefer understand and foresee price trends driven by market dynamics. Relations (1.1) use assumption that trade value 1$ has same probability as trade value of 1 B$ and that seems a little bit far from reality. The VWAP (1.2; 1.3) takes into account the economic value, the market impact of the large volume trades $U(t_i)$ with price $p(t_i)$. In simple words, description of irregular asset prices via the standard frequency price probability (1.1) is complemented or replaced by description of price dynamics via certain unit price measure $\mu$ (1.2; 1.3; 1.5). Different unit price measures can be proposed for asset price description. To simplify the paper further we call the mean, mean square, mean *n-th* degree price with respect to certain unit price measure as *n-th* statistical moments.

In this paper we discuss the economic meaning of VWAP and introduce the unit price measure that describes *n-th* price statistical moments via ratios of sum of *n-th* degrees values and volumes of trades aggregated during interval $\Delta$. Such description establishes a strong link between market trade trends and evolution of price statistical moments. The usage of unit measures allows model fluctuations, volatility and etc. similar to approach based on frequency price probability (1.1).

With respect to proposed unit price measure we introduce mean square price $p(2;t)$ as the ratio of the sum of squares of the values $C_i^2$ to the sum of squares of the volumes $U_i^2$ of all trades $i$ performed during interval $\Delta$. We give formal definitions of the mean square price $p(2;t)$ in Sec.2. It is obvious that the mean square price $p(2;t)$ and price volatility are not the only second-degree variables. All macro variables as bank rates, inflation, currency exchange rates, return on investment, GDP growth rates and etc., - fluctuate. Investment and profits, supply and demand, production function and all macroeconomic and financial variables fluctuate. We introduce mean square price $p(2;t)$ and price volatility in a way that establishes direct functional links between price volatility, market trades and macroeconomic variables. Thus price volatility determined by unit price measure depends upon properties of second-degree market trades. We show that the description and forecasting of price volatility implies modeling dynamics of the second-degree trades, economic and financial variables aggregated during interval $\Delta$. We call the description of aggregated second-degree trades as the second-order economic theory and develop the beginning of this theory.

In this paper we only start study of the second-order economic theory. We model price volatility determined by unit price measure and describe the second-degree trades collected during interval $\Delta$. We introduce main notions and derive equations that describe evolution of squares of market trades. It is generally accepted that agents perform market transactions under impact of their expectations. We describe expectations those approve the second-degree trades and derive equations that govern evolution of the second-degree expectations. We show how equations on the second-degree trades cause equations on mean square price *p(2;t)*. We believe that math complexity of our methods and equations corresponds the complexity of this economic problem.

In Sec.2 we introduce the unit price measure and describe price volatility as function of first and second-degree values and volumes of trades performed during interval $\Delta$. In Sec.3 we describe second-degree macro trades and their flows. In Sec. 4 we introduce collective expectations and collective expected trades that govern the second-degree trades. In Sec 5 we describe disturbances of the second-degree trades and model small price and volatility disturbances. Sec.6 - Conclusion. In Appendix A we derive relations on second-degree trades and their flows. In Appendix B we introduce math for collective expectations. In Appendix C we derive equations on second-degree trades. Appendix D describes perturbations of second-degree trades, price and volatility.

Equations (3.4) denote Section 3 and equation 4. We use roman letters *A, B, d* to denote scalars and bold ***B, P, v*** – to denote vectors. We assume that readers are familiar with basic notions and methods of stochastic dynamic systems and usage of probability density functions, statistical moments and characteristic functions. We refer (Klyatskin, 2005; 2015) for details. We hope that notations of partial differential equations don't need clarifications.

## 2. Price and volatility

Price is one of most common and basic notions of economics and finance. As usual common notions have numerous definitions. More then hundred years ago Frank Fetter mentioned 117 price definitions (Fetter, 1912). We take the trivial price $p$ definition also mentioned by Fetter: "Ratio-of-exchange definitions of price in terms of value in the sense of a mere ratio of exchange" – price $p$ is coefficient between the value $C$ and the volume $U$ of market trade:

$$C = pU \tag{2.1}$$

We call the pair *(U,C)* as the market trade or the transaction. To develop smooth description of irregular price (2.1) dynamics we introduce unit price measure and model mean price with respect to this measure. We study mean price, price volatility with respect to this measure and establish functional relations between unit measure and properties of market trades.

However economic agents go into market transactions under action of their personal expectations. Agents may develop their personal price expectations and believes on base of any of 117 price definitions mentioned by Fetter (1912). Agents may accept any possible price assumptions, prospects and forecasts. Agents may use any arbitrary price models and adopt reasons that support their decisions to execute the market trade with particular price value (2.1). For sure agents price expectations impact the properties of the price value (2.1). We use the results of the performed trades to describe dependence of price time-series random properties (2.1) on properties of values and volumes of performed market trades.

Relations (1.1) define the most common treatment of probability of market price. As we mentioned above (1.1) is not the only way to describe irregular trade price properties. VWAP (1.2; 1.3) defines the mean price with respect to measure $\mu$ (1.5) determined by aggregated value $C(1;t)$ and volume $U(1;t)$ of market trades during interval $\Delta(t)$ (1.4) and not by price frequencies (1.1). The economic collision between two treatments of mean price determined by price frequencies (1.1) and by unit measure $\mu$ (1.5) may be the origin of excess and unexpected losses for investors and may cause ambivalent economic and financial forecasting and projections.

Below we introduce new unit price measure $\eta$ that for all $n=1,2,..$ help describe price *n-th* statistical moments and establishes a functional link between price statistical moments and market trades. Below we use terms mean, volatility and etc., with respect to the unit price measure $\eta(p)$. Let's mention important issue: one can determine any unit measure by defining all its statistical moments. In other words – to determine unit measure $\eta(p)$ it is sufficient for all $n=1,2,\ldots$ determine price statistical moments:

$$p(n;t) = \sum \eta(p)\, p^n \quad ; \quad n = 1,2,\ldots \tag{2.2}$$

To define *n-th* price moments $p(n;t)$ let's start with *n-th* degree of a single trade that follows from (2.1):

$$C^n(t_i) = p^n(t_i) U^n(t_i) \quad ; \quad n = 1,2,\ldots \tag{2.3}$$

To define mean the *n-th* degree price moment $p(n;t)$ from (2.3) we follow the logic of VWAP (1.3) and introduce the sum of the *n-th* degree of the value $C^n(t_i)$ and of *n-th* degree of the volume $U^n(t_i)$ of all $N(t)$ trades during interval $\Delta(t)$ (1.4) at moment $t$ as:

$$C(n;t) = \sum_{i=1}^{N(t)} C^n(t_i) \quad ; \quad U(n;t) = \sum_{i=1}^{N(t)} U^n(t_i) \tag{2.4}$$

$C(n;t)$ and $U(n;t)$ (2.4) determine mean *n-th* degree price $p(n;t)$ (N-VWAP) of $N(t)$ market trades (2.3) performed during interval $\Delta(t)$ (1.4) as:

$$C(n;t) = p(n;t) U(n;t) \tag{2.5}$$

For $n=1$ relations (2.4, 2.5) define $p(1;t)$ (1.3) - volume weighted average price (VWAP) described by (Berkowitz et.al 1988; Buryak and Guo, 2014; Guéant and Royer, 2014; Busseti and Boyd, 2015; Padungsaksawasdi and Daigler, 2018; CME Group, 2020). One can present N-VWAP (2.5) similar to (1.3):

$$p(n;t) = \frac{1}{U(n;t)} \sum_{i=1}^{N(t)} p^n(t_i)\, U^n(t_i) = \frac{C(n;t)}{U(n;t)} \tag{2.6}$$

We underline that VWAP *p(1;t)* (1.2) and N-VWAP *p(n;t)* (2.5; 2.6) describe price statistical moments determined by the sum of *n-th* degree of value *C(n;t)* and the *n-th* degree of volume *U(n;t)* of market trades aggregated during interval *Δ(t)* (1.4). Relations (2.6) show that price statistical moments *p(n;t)* are obtained by weighting the *n-th* degree price $p^n(t_i)$ by the *n-th* degree volume $U^n(t_i)$ of market trades.

One-moment price statistical moments *p(n;t)* (2.5; 2.6) describe the unit measure *η(p;t)* and (2.5; 2.6) determine characteristic function $F_p(x;t)$ (2.7-2.9) that completely describes the unit price measure *η(p;t)*:

$$F_p(x;t) = \sum_{i=1}^{\infty} \frac{i^n}{n!} p(n;t)\, x^n \tag{2.7}$$

$$F_p(x;t) = \int dp\, \eta(p;t) \exp(ixp) \tag{2.8}$$

$$\frac{d^n}{(i)^n dx^n} F_p(x;t)|_{x=0} = \int dp\, \eta(p;t) \mathrm{p}^n = p(n;t) \tag{2.9}$$

One can consider random price *p(t)* as a random process. If so, the measure *η(p;t)* or price random process should be described by characteristic functional that is alike to (2.8), but we omit here the definition of price characteristic functional for simplicity. We refer (Klyatskin, 2005; 2015) as helpful source for usage of the characteristic functional and functional calculus for description of random processes and stochastic dynamic systems.

Relations (2.5; 2.6) for first two price statistical moments define price volatility $\sigma_p^2(t)$ with respect to the unit price measure *η(p;t)* (2.7; 2.8) during interval *Δ(t)* as a difference between mean squares *p(2;t)* and squares of mean price $p^2(1;t)$:

$$\sigma_p^2(t) = p(2;t) - p^2(1,t) = \frac{C(2;t)}{U(2;t)} - \frac{C^2(1;t)}{U^2(1;t)} \tag{2.10}$$

Relations (2.10) present price volatility $\sigma_p^2(t)$ with respect to unit measure *η(p;t)* as function of *C(n;t)* and *U(n;t)* (2.4) for *n=1,2*. To forecast the price volatility (2.10) with respect to measure *η(p;t)* (2.7-2.9) one should model dynamics of *p(1;t)* and *p(2;t)* (2.6). Description of *p(1;t)* relates to the first-order economic theory and models the evolution of macroeconomic variables under action of market trades (Olkhov, 2018-2019b). To model *p(2;t)* one should develop description of second-degree trades (2.4) for *n=2*. Direct functional link between price volatility $\sigma_p^2(t)$ with respect to the unit measure *η(p;t)* (2.7-2.9) and market trades (2.4) for *n=1,2* reflects existing constraints between market dynamics and price volatility.

Current macroeconomic and financial models describe mutual dependence of macroeconomic and financial variables of the first-degree. Macro variables are composed as sum of corresponding variables of all economic agents in economy during interval $\Delta$. For example, macro investment is determined as investment made by all agents in economy during interval $\Delta$. Macro credit, taxes, consumption, output, profits and etc., are determined as sum (without doubling) of credits, taxes, consumption, output, profits and etc., of all agents in economy during interval $\Delta$. Relations (2.10) show that volatility with respect to the unit measure $\eta(p;t)$ depends on second price statistical moment $p(2;t)$ determined by (2.4-2.7) for $n=2$. Hence description and forecasting of price volatility requires modeling the second-degree trades (2.4) aggregated during time interval $\Delta$. It is obvious that second-degree trades (2.4) impact second-degree economic and financial variables involved into transactions. In particular, the second-degree value (2.4) impact second-degree funds of agents involved into transactions. Second-degree volumes (2.4) impact second-degree goods commodities that belong to agents involved into transactions. Thus description of second-degree trades (2.4) establishes ground for modeling second-degree macroeconomic variables and their volatility.

In the next section consider description of the second-degree macro trades.

## 3. Second-degree trades

We describe the second-degree macro trades using the methods (Olkhov, 2016a-2020) and refer these papers for details. For reader's convenience we present brief consideration of our approach. We develop the theory of the second-degree macro trades on several assumptions.

First, we assume that it is possible develop and use the unified methodology for agents risk assessments based on numerical continuous risk scoring. Up now risk ratings are noted by letters. Each agency - S&P, (2014, 2016), Moody's, (2010, 2018), Fitch, (2018) - introduces its own letter-ratings methodology. However numerical credit scoring was suggested at least eighty years ago by Durand (1941) and then sixty years ago by Myers and Forgy (1963). Last years researchers use numerical risk rating to compare risk assessment provided by different rating agencies (Beaver et.al., 2006; Poon and Shen, 2020). We assume that no internal econometric obstacles may prevent the transfer from letters to numerical risk scoring. This transition depends mostly on the business interests of major ratings companies. Development of unified methodology that can be adopted by major business players should deliver benefits to rating agencies, investors, financial markets, business and economic authorities. Unified numerical continuous risk grades methodology will significantly improve methods for modeling and forecasting macroeconomic and financial processes. Introduction of continuous risk grades is a natural development of numerical grades methodology. We don't specify any particular risks like credit or inflation risk but regard all possible economic, financial, political and etc., risks those impact economic and financial development.
Second, we assume that it is possible to assess risk ratings for almost all economic agents like international banks, corporations, local firms, companies and even households. That will distribute all economic agents by their numerical risk ratings as coordinates of the risk space.
Third, agents in the economy always act under pressure of numerous risks. Assessment of agents ratings for one risk distribute agents over *1*-dimensional risk space. Simultaneous assessment of agents ratings for two or three risks distribute agents by two or three dimensional risk space. Absolute numerical values of risk ratings have no sense. For certainty, for any risk one may take most secure rating to be equal zero and the most risky rating to be equal 1. Thus agents continuous numerical ratings for a single risk fill unit interval [0,1] and agents ratings for $m$ risks fill unit cube in space $R^m$.
Forth. Explicit description of numerous separate agents of the entire economy is almost impossible. It requires too much exact data of economic and financial variables for millions of separate agents, data about all particular market trades between agents, risk assessment of all agents and etc. It is almost impossible to collect precise data for all agents and small perturbations could move the forecasts in the wrong directions. To solve the problem we reduce the accuracy of economic description. We aggregate description of agents in the risk economic space and derive the continuous media approximation (Olkhov, 2016a, 2018; 2019a -2019c) that describes collective transactions. Below we explain meaning of the continuous media approximation for transactions in more detail.
Let's explain the benefits of our approach for economic modeling. First, transition from letters to continuous numerical risk scoring uncovers the hidden motion of economic agents. Indeed, rating agencies for decades use transition matrix to describe probability of agent's transfer from particular risk grade to a different one (Belkin, 1998; Schuermann and Jafry, 2003; Ho et.al, 2017; S&P, 2018). Introduction of numerical continuous scoring defines the numerical distance between risk ratings and hence introduces the mean velocity of agent's motion during risk transition. The motion of agents in the risk economic space defines flows of macro variables, trades and expectations carried by economic agents within risk motion. Below we use this approach to develop description of second-degree trades and refer (Olkhov, 2018, 2019b, 2019c) for all further details and definitions.

### *3.1. Second-degree transactions*

To start with let's regard economy as a system of numerous economic agents those perform various economic and financial transactions. We assume that agents perform market trades under action of $m$ risks and it is possible to assess risk ratings for all agents in the economy. Each economic agent $i$ is determined by its risk coordinates - numerical continuous risk ratings $\boldsymbol{x}$

$$\boldsymbol{x} = (x_1, \ldots x_m) \;\; ; \;\; 0 \le x_i \le 1 \;\; ; \;\; i = 1,..m \tag{3.1}$$

Coordinates $\boldsymbol{x}$ of agents fill unit economic domain (3.1) of risk space $R^m$. Further we note risk space as economic space. We treat risk assessments of agents as a procedures similar to measurements of coordinates of physical particles and call agents numerical continuous risk ratings $\boldsymbol{x}$ (3.1) as agents coordinates in economic domain (3.2):

$$0 \le x_i \le 1 \;\; ; \;\; i = 1,..m \tag{3.2}$$

Let's take that agent $i$ with risk coordinates $\boldsymbol{x}$ executes transaction with agent $j$ with risk coordinates $\boldsymbol{y}$. To simplify the model we assume that at moment $t$ agent $i$ at point $\boldsymbol{x}$ sells the volume $U_{ij}$ of variable $E$ to agent $j$ at point $\boldsymbol{y}$. As $E$ one can consider any assets, commodities, goods, currency, service, credits, investment and etc. Let's assume that the volume $U_{ij}$ of this particular transaction between agents $i$ and $j$ has the value $C_{ij}$. We define the market trade of the first-degree at moment $t$ between agents $i$ and $j$ as two component function $\boldsymbol{b}_{ij}$ :

$$\boldsymbol{b}_{ij}(1; t, \boldsymbol{z}) = \left(U_{ij}(t, \boldsymbol{z}); C_{ij}(t, \boldsymbol{z})\right) \; ; \; \boldsymbol{z} = (\boldsymbol{x}, \boldsymbol{y}) \tag{3.3}$$

Coordinates $\boldsymbol{x}$ and $\boldsymbol{y}$ of agents $i$ and $j$ involved into this trade establish economic domain with double dimension (3.1; 3.2; 3.4; 3.5) with coordinates $z=(\boldsymbol{x},\boldsymbol{y})$ :

$$\boldsymbol{z} = (\boldsymbol{x}, \boldsymbol{y}) \;\; ; \;\; \boldsymbol{x} = (x_1 \ldots x_m) \;\; ; \;\; \boldsymbol{y} = (y_1 \ldots y_m) \tag{3.4}$$

$$0 \le x_i \le 1 \;\; ; \; 0 \le y_j \le 1 \; ; \;\; i = 1,..m \; ; \;\; j = 1,..m \tag{3.5}$$

We define the price $p_{ij}$ of the trade $\boldsymbol{b}_{ij}$ (3.3) between agents $i$ and $j$ similar to (2.1):

$$C_{ij}(t, \boldsymbol{z}) = p_{ij}(t, \boldsymbol{z}) U_{ij}(t, \boldsymbol{z}) \tag{3.6}$$

Let's assume that the execution of the trade $\boldsymbol{b}_{ij}$ (3.3) takes certain time $dt$. The volume $U_{ij}$ of the trade $\boldsymbol{b}_{ij}$ (3.3) change the amount of economic or financial variable $E$ of agents $i$ and $j$ involved into the trade during time $dt$ and the value $C_{ij}$ change the amount of their funds. We consider the trades as the rate of change of economic and financial variables of economic agents. Let's define the second-degree trade $\boldsymbol{b}_{ij}(2;t,z)$ as:

$$\boldsymbol{b}_{ij}(2; t, \boldsymbol{z}) = \left(U_{ij}^2(t, \boldsymbol{z}); C_{ij}^2(t, \boldsymbol{z})\right) \tag{3.7}$$

We define transactions $\boldsymbol{b}_{ij}(n;t,z)$ of degree $n$ as

$$\boldsymbol{b}_{ij}(n; t, \boldsymbol{z}) = \left(U_{ij}^n(t, \boldsymbol{z}); C_{ij}^n(t, \boldsymbol{z})\right) \tag{3.8}$$

We determine the *n-th* degree price $p^n_{ij}(t,z)$ of trade $\boldsymbol{b}_{ij}(n;t,z)$ (3.8) similar to (2.3) as:

$$C^n_{ij}(t,\boldsymbol{z}) = p^n_{ij}(t,\boldsymbol{z})U^n_{ij}(t,\boldsymbol{z}) \tag{3.9}$$

To avoid excess accuracy and to derive the economic continuous media approximation for the second-degree trades we replace description of the second-degree trades $\boldsymbol{b}_{ij}(2;t,z)$ between individual agents $i$ at $\boldsymbol{x}$ and $j$ at $\boldsymbol{y}$ by description of collective second-degree trades $\boldsymbol{B}(2;t,z)$, $z=(\boldsymbol{x},\boldsymbol{y})$ between points $\boldsymbol{x}$ and $\boldsymbol{y}$. To do that we introduce small scale $d$ and a unit volume $dV(z)$ in economic domain (3.4, 3.5)

$$dV(\boldsymbol{z}) = dV(\boldsymbol{x})dV(\boldsymbol{y})\ ;\ dV(\boldsymbol{x}) = d^n\ ;\ dV(\boldsymbol{y}) = d^n\ ;\ \boldsymbol{z} = (\boldsymbol{x},\boldsymbol{y}) \tag{3.10}$$

We assume that scale $d << 1$ but each unit volume $dV(\boldsymbol{x})$ and $dV(\boldsymbol{y})$ contains a lot of agents with risk coordinates inside $dV(\boldsymbol{x})$ and $dV(\boldsymbol{y})$. Let's assume that during time interval $\Delta$ agents inside $dV(\boldsymbol{x})$ and $dV(\boldsymbol{y})$ perform a lot of mutual transactions. We define the collective second-degree trade $\boldsymbol{B}(2;t,z)$ between points $\boldsymbol{x}$ and $\boldsymbol{y}$ as a sum of all second-degree trades $\boldsymbol{b}_{ij}(2;t,z)$ of agents $i$ with coordinates in a unit volume $dV(\boldsymbol{x})$ and agents $j$ with coordinates in a unit volume $dV(\boldsymbol{y})$ (3.10) and then average this sum during the interval $\Delta$ as:

$$\boldsymbol{B}(2;t,\boldsymbol{z}) = \sum_{i\in dV(\boldsymbol{x});j\in dV(\boldsymbol{y});\,\Delta} \boldsymbol{b}_{i,j}(2;t,\boldsymbol{z}) \tag{3.11}$$

$$\sum_{i\in dV(\boldsymbol{x});j\in dV(\boldsymbol{y});\,\Delta} \boldsymbol{b}_{i,j}(2;t,\boldsymbol{z}) = \frac{1}{\Delta}\int_{t-\Delta/2}^{t+\Delta/2} d\tau\ \sum_{i\in dV(\boldsymbol{x});j\in dV(\boldsymbol{y})} \boldsymbol{b}_{i,j}(2;\tau,\boldsymbol{z}) \tag{3.12}$$

$$\boldsymbol{B}(2;t,\boldsymbol{z}) = \big(U(2;t,\boldsymbol{z});C(2;t,\boldsymbol{z})\big)\ ;\ \boldsymbol{z} = (\boldsymbol{x},\boldsymbol{y}) \tag{3.13}$$

$$U(2;t,\boldsymbol{z}) = \sum_{i\in dV(\boldsymbol{x});j\in dV(\boldsymbol{y});\,\Delta} U^2_{ij}(t,\boldsymbol{z}) \tag{3.14}$$

$$C(2;t,\boldsymbol{z}) = \sum_{i\in dV(\boldsymbol{x});j\in dV(\boldsymbol{y});\,\Delta} C^2_{ij}(t,\boldsymbol{z}) \tag{3.15}$$

The second-degree collective trades $\boldsymbol{B}(2;t,z)$ (3.11-3.15) define the second-degree price $p(2;t,z)$ of trades between agents at points $\boldsymbol{x}$ and agents at point $\boldsymbol{y}$ in domain (3.4, 3.5) averaged during interval $\Delta$ similar to (2.5):

$$C(2;t,\boldsymbol{z}) = p(2;t,\boldsymbol{z})U(2;t,\boldsymbol{z}) \tag{3.16}$$

Relation (3.16) describes mean second-degree price $p(2;t,z)$ between points $\boldsymbol{x}$ and $\boldsymbol{y}$ averaged during interval $\Delta$. Relation (2.5; 2.7) for $n=2$ describe mean second-degree price $p(2;t)$ of all trades in the economy during interval $\Delta$. To derive (2.5; 2.7) one should take integral by collective second-degree trades $\boldsymbol{B}(2;t,z)$ over economic domain (3.4, 3.5) or calculate the sum over all agents in economy involved into trades during time interval $\Delta$:

$$\boldsymbol{B}(2;t) = \big(U(2;t);C(2;t)\big) \tag{3.17}$$

$$U(2;t) = \int d\boldsymbol{z}\, U(2;t,\boldsymbol{z})\quad ;\quad C(2;t) = \int d\boldsymbol{z}\, C(2;t,\boldsymbol{z}) \tag{3.18}$$

$$C(2;t) = p(2;t)U(2;t) \tag{3.19}$$

Relations (3.19) define price $p(2;t)$ in the same way as (2.5) for $n=2$. Description of price (2.5; 2.7) requires models of the second-degree trades $\boldsymbol{B}(2;t)$ and $\boldsymbol{B}(2;t,z)$.

### *3.2. Flows of the second-degree transactions*

Let's start with the definition of agent's velocities in the economic domain (3.1, 3.2). Any economic activity of agents is the source of risks and agents always act under the pressure of risks – credit, inflation, market, political and etc. Rating agencies for decades provide assessments of risk transition matrices (Belkin, 1998; Schuermann and Jafry, 2003; Ho et.al, 2017; S&P, 2018). Up now risk grades are determined by letters and elements $a_{ij}$ of transition matrices define probabilities of transition from grade $i$ to $j$ during certain time interval $T$. As usual interval $T$ equals half year, one, two three years. If and when ratings methodology uses numerical continuous grades then transition matrices get completely different and much more important and reasonable economic meaning. For numerical continuous grades the transition from rating $x_i$ to $x_j$ defines numerical interval $l_{ij}$ :

$$l_{ij} = x_j - x_i \tag{3.20}$$

Transition from $x_i$ to $x_j$ takes time T. Hence element $a_{ij}$ of transition matrices defines the probability of agent's motion in the economic domain (3.2) during time $T$ with velocity $v_{ij}$ :

$$v_{ij} = \frac{l_{ij}}{T} \; with\, probability\; a_{ij}; \quad \sum_j a_{ij} = 1 \tag{3.21}$$

As we mentioned above, introduction of agents velocities in the economic domain is important contribution of the transfer from letters to continuous numerical risk grades. Hence transition matrices define mean velocity of agent in point $x_i$ during time interval $T$ as:

$$v(t, x_i) = \sum_{j=1}^{K} v_{ij} a_{ij} = \frac{1}{T} \sum_{j=1}^{K} l_{ij} a_{ij} \tag{3.22}$$

Here $K$ means the number of different numerical risk grades that defines the degree $K$x$K$ of the transition matrix. Thus the transition matrices can define mean velocities of agents in the economic domain. Risk motion of agents induces flows of economic and financial variables carried by agents in risk domain and flows of market trades as well (Olkhov, 2018-2019c). Indeed, each agent in economic domain with velocity $\boldsymbol{v}$ carries its economic and financial variables. Collective effect of such transport of economic variables by all agents in a small volume $dV(\boldsymbol{x})$ is described by flow of variables in (3.2). Similar effect defines the flow of collective trades in economic domain with double dimension (3.4,3.5).

To describe the flows of the second-degree transactions $\boldsymbol{B}(2;t,z)$ let's assume that in economic domain (3.2) agent $i$ at a moment $t$ have risk coordinates $\boldsymbol{x}=(x_1,...x_m)$ and velocities $\boldsymbol{v_x}=(v_{x1},...v_{xm})$. Let's take that agent $i$ at point $\boldsymbol{x}$ executes transaction $\boldsymbol{b}_{ij}(2;t,z)$ with agent $j$ at point $\boldsymbol{y}=(y_1,\ldots y_m)$ and velocities $\boldsymbol{v}_y=(v_{y1},\ldots v_{ym})$. Similar to (Olkhov, 2018; 2019a – 2019c) let's define the flows $\boldsymbol{p}_{ij}(2;t,\boldsymbol{z})$ of the transactions $\boldsymbol{b}_{ij}(2;t,\boldsymbol{z})$ (3.7) between agents $i$ and $j$ as:

$$\boldsymbol{p}_{ij}(2;t,\boldsymbol{z}) = \left(\boldsymbol{p}_{Uij}(2;t,\boldsymbol{z}), \boldsymbol{p}_{Cij}(2;t,\boldsymbol{z})\right) \quad (3.23)$$

$$\boldsymbol{p}_{Uij}(2;t,\boldsymbol{z}) = \left(\boldsymbol{p}_{Uxij}(2;t,\boldsymbol{z});\ \boldsymbol{p}_{Uyij}(2;t,\boldsymbol{z})\right) \ ; \ \boldsymbol{p}_{Cij}(2;t,\boldsymbol{z}) = \left(\boldsymbol{p}_{Cxij}(2;t,\boldsymbol{z});\ \boldsymbol{p}_{Cyij}(2;t,\boldsymbol{z})\right) \quad (3.24)$$

$$\boldsymbol{p}_{Uxij}(2;t,\boldsymbol{z}) = U_{ij}^2(t,\boldsymbol{z})\boldsymbol{v}_{xi}(t,\boldsymbol{x}) \quad ; \quad \boldsymbol{p}_{Uyij}(t,\boldsymbol{z}) = U_{ij}^2(t,\boldsymbol{z})\boldsymbol{v}_{yj}(t,\boldsymbol{y}) \quad (3.25)$$

$$\boldsymbol{p}_{Cxij}(t,\boldsymbol{z}) = C_{ij}^2(t,\boldsymbol{z})\boldsymbol{v}_{xi}(t,\boldsymbol{x}) \quad ; \quad \boldsymbol{p}_{Cijy}(t,\boldsymbol{z}) = C_{ij}^2(t,\boldsymbol{z})\boldsymbol{v}_{yj}(t,\boldsymbol{y}) \quad (3.26)$$

The flows $\boldsymbol{p}_{ij}(2;t,\boldsymbol{z})$ (3.23) denote the flows $\boldsymbol{p}_{Uij}(2;t,\boldsymbol{z})$ (3.25) that carry the square of volume $U_{ij}^2$ of the trade $\boldsymbol{b}_{ij}(2;t,\boldsymbol{z})$ (3.7). The flows $\boldsymbol{p}_{Cij}(t,\boldsymbol{z})$ (3.26) carry the square of the value $C_{ij}^2$ of the trade $\boldsymbol{b}_{ij}(2;t,\boldsymbol{z})$ (3.7). We determine the cumulative flows $\boldsymbol{P}(2;t,\boldsymbol{z})$ of the macro second-degree transaction $\boldsymbol{B}(2;t,\boldsymbol{z})$, $\boldsymbol{z}=(\boldsymbol{x},\boldsymbol{y})$ between points $\boldsymbol{x}$ and $\boldsymbol{y}$ similar to (3.11-3.15) as aggregation of flows $\boldsymbol{p}_{ij}(2;t,\boldsymbol{z})$ (3.23) of all second-degree transactions $\boldsymbol{b}_{ij}(2;t,\boldsymbol{z})$ (3.7) between agents in small volumes $dV(\boldsymbol{x})$ and $dV(\boldsymbol{y})$ (3.10) in economic domain (3.4, 3.5) and averaging during interval $\Delta$. Due to (3.11-3.15) we introduce flows $\boldsymbol{P}(2;t,\boldsymbol{z})$ and velocities $\boldsymbol{v}(2;t,\boldsymbol{z})$ as:

$$\boldsymbol{P}(2;t,\boldsymbol{z}) = \left(\boldsymbol{P}_U(2;t,\boldsymbol{z}), \boldsymbol{P}_C(2;t,\boldsymbol{z})\right) \ ; \ \boldsymbol{z} = (\boldsymbol{x},\boldsymbol{y}) \quad (3.27)$$

$$\boldsymbol{P}_U(2;t,\boldsymbol{z}) = \sum_{i\in dV(\boldsymbol{x});j\in dV(\boldsymbol{y})}{}_{\Delta}\, \boldsymbol{p}_{Uij}(2;t,\boldsymbol{z}) \quad (3.28)$$

$$\boldsymbol{P}_C(t,\boldsymbol{z}) = \sum_{i\in dV(\boldsymbol{x});j\in dV(\boldsymbol{y})}{}_{\Delta}\, \boldsymbol{p}_{Cij}(2;t,\boldsymbol{z}) \quad (3.29)$$

$$\boldsymbol{P}_U(2;t,\boldsymbol{z}) = \left(\boldsymbol{P}_{Ux}(t,\boldsymbol{z}); \boldsymbol{P}_{Uy}(t,\boldsymbol{z})\right) \ ; \ \boldsymbol{P}_C(t,\boldsymbol{z}) = \left(\boldsymbol{P}_{Cx}(t,\boldsymbol{z}); \boldsymbol{P}_{Cy}(t,\boldsymbol{z})\right) \quad (3.30)$$

$$\boldsymbol{P}_{Ux}(2;t,\boldsymbol{z}) = \sum_{i\in dV(\boldsymbol{x});j\in dV(\boldsymbol{y})}{}_{\Delta}\, U_{ij}^2(t,\boldsymbol{z})\boldsymbol{v}_i(t,\boldsymbol{x}) = U(2;t,\boldsymbol{z})\boldsymbol{v}_{Ux}(t,\boldsymbol{z}) \quad (3.31)$$

$$\boldsymbol{P}_{Uy}(2;t,\boldsymbol{z}) = \sum_{i\in dV(\boldsymbol{x});j\in dV(\boldsymbol{y})}{}_{\Delta}\, U_{ij}^2(t,\boldsymbol{z})\boldsymbol{v}_j(t,\boldsymbol{y}) = U(2;t,\boldsymbol{z})\boldsymbol{v}_{Uy}(t,\boldsymbol{z}) \quad (3.32)$$

$$\boldsymbol{P}_{Cx}(2;t,\boldsymbol{z}) = \sum_{i\in dV(\boldsymbol{x});j\in dV(\boldsymbol{y})}{}_{\Delta}\, C_{ij}^2(t,\boldsymbol{z})\boldsymbol{v}_i(t,\boldsymbol{x}) = C(2;t,\boldsymbol{z})\boldsymbol{v}_{Cx}(t,\boldsymbol{z}) \quad (3.33)$$

$$\boldsymbol{P}_{Cy}(2;t,\boldsymbol{z}) = \sum_{i\in dV(\boldsymbol{x});j\in dV(\boldsymbol{y})}{}_{\Delta}\, C_{ij}^2(t,\boldsymbol{z})\boldsymbol{v}_j(t,\boldsymbol{y}) = C(2;t,\boldsymbol{z})\boldsymbol{v}_{Cy}(t,\boldsymbol{z}) \quad (3.34)$$

$$\boldsymbol{v}(2;t,\boldsymbol{z}) = \left(\boldsymbol{v}_U(2;t,\boldsymbol{z});\ \boldsymbol{v}_C(2;t,\boldsymbol{z})\right) \quad (3.35)$$

$$\boldsymbol{v}_U(2;t,\boldsymbol{z}) = \left(\boldsymbol{v}_{Ux}(2;t,\boldsymbol{z})\,; \boldsymbol{v}_{Uy}(2;t,\boldsymbol{z})\right) \ ; \ \boldsymbol{v}_C(2;t,\boldsymbol{z}) = \left(\boldsymbol{v}_{Cx}(2;t,\boldsymbol{z})\,; \boldsymbol{v}_{Cy}(2;t,\boldsymbol{z})\right) \quad (3.36)$$

The flows $\boldsymbol{P}(2;t,\boldsymbol{z})$ (3.27-3.34) of the second-degree transaction $\boldsymbol{B}(2;t,\boldsymbol{z})$ between points $\boldsymbol{x}$ and $\boldsymbol{y}$ describe the amounts of second-degree volume $U(2;t,\boldsymbol{z})$ (3.14) and the value $C(2;t,\boldsymbol{z})$ (3.15) of trades $\boldsymbol{B}(2;t,\boldsymbol{z})$ (3.11) carried by transactions velocities $\boldsymbol{v}(2;t,\boldsymbol{z})$ (3.35, 3.36) through $2m$-dimensional economic domain (3.4, 3.5). Velocity $\boldsymbol{v}_U(2;t,\boldsymbol{z})$ (3.36) defines the motion of second-degree volume $U(2;t,\boldsymbol{z})$ (3.14) and may be different from velocity $\boldsymbol{v}_C(2;t,\boldsymbol{z})$ (3.36) that describes the motion of second-degree value $C(2;t,\boldsymbol{z})$ (3.15).

Integral of transactions $\boldsymbol{B}(2;t,\boldsymbol{z})$ and their flows $\boldsymbol{P}(2;t,\boldsymbol{z})$ by $d\boldsymbol{y}$ over entire economic domain (3.4, 3.5) defines the second-degree sell-transactions $\boldsymbol{B}_s(2;t,\boldsymbol{x})$ from point $\boldsymbol{x}$. Integral by $d\boldsymbol{x}$ defines buy-transactions $\boldsymbol{B}_s(2;t,\boldsymbol{y})$ at point $\boldsymbol{y}$. Integral by $dz=dxdy$ defines total transactions $\boldsymbol{B}(2;t)$ in the entire economy at moment $t$ and their flows (see Appendix A). Long definitions (Appendix A) demonstrate the hidden complexities of the second-degree economic processes in economy and are useful for description of macroeconomic evolution.

Introduction of the second-degree transactions and their flows as functions of risk in the economic domain (3.4,3.5) outlines contribution of our approach to the price volatility modeling, macroeconomics and finance. To describe and forecast price volatility (2.10) one should model the second-degree transactions.

The second-degree transactions define different forms of the price: (3.6, 3.9) define price $p_{ij}(t,z)$ and $p_{ij}^n(t,z)$ of particular transaction between agents $i$ and $j$ at points $\boldsymbol{x}$ and $\boldsymbol{y}$, $z=(\boldsymbol{x},\boldsymbol{y})$. Relations (3.16) define average square price $p(2;t,z)$ at moment $t$ of all transactions between agents in small volumes near points $\boldsymbol{x}$ and $\boldsymbol{y}$, averaged during interval $\Delta$. Relations (3.19) introduce average squared price $p(2;t)$ of all transactions in the economy averaged during time interval $\Delta$ similar to (2.5; 2.7) for $n=2$. Relations (A.1-A.4) in Appendix A introduce average square price $p_s(2;t,\boldsymbol{x})$ for all sell-trades from point $\boldsymbol{x}$ and average square price $p_b(2;t,\boldsymbol{y})$ for all buy-trades at point $\boldsymbol{y}$ at moment $t$ averaged during interval $\Delta$ as:

$$C_s(2;t,\boldsymbol{x}) = p_s(2;t,\boldsymbol{x})U_s(2;t,\boldsymbol{x}) \tag{3.37}$$

$$C_b(2;t,\boldsymbol{y}) = p_b(2;t,\boldsymbol{y})U_b(2;t,\boldsymbol{y}) \tag{3.38}$$

Relations (3.6, 3.9, 3.16, 3.19, 3.37, 3.38) describe different forms of price. Key contribution of our approach to macroeconomic modeling: evolution of transactions depends on transactions flows. Description of price must take into account transactions flows – these new economic factors impact price dynamics. Below we derive equations that govern the second-degree transactions and their flows.

## 4. Collective expectations

It is generally considered that agents take trade decisions under their expectations. Expectations play the crucial role for the market trade performance. Decades of research describe impact of expectations on financial markets, price dynamics and economic evolution (Muth, 1961; Lucas, 1972; Blume and Easley, 1984; Brunnermeier and Parker, 2005; Dominitz and Greenwood and Shleifer, 2014). Observations and measurements of expectations is a difficult problem (Manski, 2004; Dominitz and Manski, 2005; Janžek and Ziherl, 2013; Manski, 2017). A lot of efforts are spent to observe, identify, estimate and measure the expectations. These problems are very complex. We suggest simplify the description of expectations and their impact on performance of market trades. To do that we regard expectations as agents assumptions and forecasts of numerous economic and financial variables, price and inflation trends, market trade activity, expectations of other agents, forecasts of any factors that can impact economy and etc. We assume that agents execute transactions under action of multiple expectations. Each agent may have multiple expectations and can perform similar transaction under different expectations. Below we model action of multiple expectations on market trades.

It seems that our assumption on multiplicity of expectations increases the complexity of the model: the difficulties of identification and measurement of expectations are increased by huge amount of different expectations of each agent. To overcome that we aggregate expectations of different agents alike to aggregation of transactions (3.11-3.15) performed by agents with risk coordinates inside the unit volume $dV(z)$ (3.10) in economic domain (3.4, 3.5). We rough the description of expectations of separate agents and introduce smooth collective expectations. This approach to expectations modeling is similar to description of macro variables and transactions developed in (Olkhov, 2016a-2019c) via the continuous economic approximation.

Multiplicity of expectations arises the problem of their comparative effect on market. Agents expectations differ from agents economic and financial variables. Agents variables like investment, assets, consumption – are additive. Sum of assets of group of agents (without overlapping) define assets of the group of agents. But expectations are not additive. Sum of expectations have no economic sense. Economic impact of the expectation that approve 1 bln.$ deal must be a little more valuable than impact of the expectation that approve 1$ deal. To collect agents expectations those approve the second-degree trades (3.7) one must take into account the economic value of the trades approved by these expectations. One can't simply sum agent's expectations to derive collective expectations of numerous agents. We state that market value of the collective expectation with respect to definite second-degree transaction should be proportional to the second-degree transaction (3.7) made under selected expectations. We assume that agents take decisions on squares of the volume $U^2$ and the value $C^2$ of the transaction under different expectations. Thus expectation that approve $U^2$ of the second-degree transaction (3.7) should be weighted by square volume $U^2$. Expectations those approve the $C^2$ of the second-degree transaction (3.7) must be weighted by square value $C^2$. We refer to (Olkhov, 2019b; 2019c) for details and apply this method to description of collective expectations of the second-degree transactions $\boldsymbol{b}_{ij}(2;t,z)$ (3.7). Let's describe collective expectations in a more formal manner.

Let's assume that agents in economy have $j=1,...K$ different expectations. Each agent involved into transaction $\boldsymbol{b}_{ij}(2;t,z)$ (3.7) between agents $i$ and $j$ should take decisions on the volume $U_{ij}$ and the value $C_{ij}$. Thus any trade is performed under action of four expectation – two expectations of agent $i$ approve the volume $U_{ij}$ and the value $C_{ij}$ and two expectations of agent $j$. Let's propose that agent $i$ take decision on the volume $U_{ij}$ under expectation $ex_{Ui}(k;t,\boldsymbol{x})$ of type $k$, and the value $C_{ij}$ under expectation $ex_{Ci}(l;t,\boldsymbol{x})$ of type $l$ , $k$, $l=1,..K$. We propose that expectations $ex_{Ui}(k;t,\boldsymbol{x})$ that approve the volume $U_{ij}(k;t,z)$ of trade may depend on expectations $ex_{Ci}(l;t,\boldsymbol{x})$ that approve the value $C_{ij}(l;t,z)$ and vice versa. Hence we propose that the volume $U_{ij}$ and the value $C_{ij}$ of the trade performed by the seller can depend on both expectations $k$ and $l$: $ex_{Ui}(k,l;t,\boldsymbol{x})$ and $ex_{Ci}(k,l;t,\boldsymbol{x})$. Let's call as second-degree trade $\boldsymbol{b}_{ij}(2;\boldsymbol{k};t,z)$ the pair $(U_{ij}^2, C_{ij}^2)$ of squares of the volume and the value made under sellers expectations of type $\boldsymbol{k}=(k,l)$ from seller $i$ at point $\boldsymbol{x}$ with buyer $j$ at point $\boldsymbol{y}$ as:

$$\boldsymbol{bs}_{ij}(2;\boldsymbol{k};t,\boldsymbol{z}) = \left(U_{ij}^2(\boldsymbol{k};t,\boldsymbol{z}); C_{ij}^2(\boldsymbol{k};t,\boldsymbol{z})\right) ; \boldsymbol{k} = (k,l) ; \ k,l = 1,...K; \ \boldsymbol{z} = (\boldsymbol{x},\boldsymbol{y}) \quad (4.1)$$

and expectations $\boldsymbol{ex}(\boldsymbol{k};t,\boldsymbol{x})$ those approve transactions (4.1) :

$$\boldsymbol{ex}(\boldsymbol{k};t,\boldsymbol{x}) = \left(ex_{Ui}(\boldsymbol{k};t,\boldsymbol{x}); ex_{Ci}(\boldsymbol{k};t,\boldsymbol{x})\right) \quad (4.2)$$

We propose that all expectations $\boldsymbol{ex}(\boldsymbol{k};t,\boldsymbol{x})$ have same measure and sum of two expectations has meaning of expectation. Thus we propose that one can sum different expectations and derive sum of expectations weighted by the volume or the value. Relations (4.1, 4.2) define the second-degree transactions and expectations of individual agent $i$ as a seller at point x. Similar relations define the second-degree transactions $\boldsymbol{b}_{ij}(2;t,z;\boldsymbol{l})$ and expectations $\boldsymbol{ex}(t,\boldsymbol{y};\boldsymbol{l})$ of a buyer at point $\boldsymbol{y}$.

$$\boldsymbol{bs}_{ij}(2;t,\boldsymbol{z};\boldsymbol{l}) = \left(U_{ij}^2(t,\boldsymbol{z};\boldsymbol{l}); C_{ij}^2(t,\boldsymbol{z};\boldsymbol{l})\right) ; \boldsymbol{l} = (k,l) ; \ k,l = 1,...K; \ \boldsymbol{z} = (\boldsymbol{x},\boldsymbol{y}) \quad (4.3)$$

and expectations $\boldsymbol{ex}(t,\boldsymbol{y};\boldsymbol{l})$ those approve transactions (4.2) :

$$\boldsymbol{ex}(t,\boldsymbol{y};\boldsymbol{l}) = \left(ex_{Ui}(t,\boldsymbol{y};\boldsymbol{l}); ex_{Ci}(t,\boldsymbol{y};\boldsymbol{l})\right) ; \boldsymbol{l} = (k,l) ; \ k,l = 1,...K \quad (4.4)$$

To derive collected expectations of agents with coordinates inside (3.10) we define new additive factor - expected trades. We weight sellers $\boldsymbol{ex}(\boldsymbol{k};t,\boldsymbol{x})$ (4.2) and buyers $\boldsymbol{ex}(t,\boldsymbol{y},\boldsymbol{l})$ (4.4) expectations by squares of the volumes and squares of the values of trades and define the second-degree sellers expected trades $\boldsymbol{et}_{ij}(\boldsymbol{k};t,z)$ as:

$$\boldsymbol{et}_{ij}(\boldsymbol{k};t,\boldsymbol{z}) = \left(et_{Uij}(\boldsymbol{k};t,\boldsymbol{z}) ; \ et_{Cij}(\boldsymbol{k};t,\boldsymbol{z})\right) \quad (4.5)$$

$$et_{Uij}(\boldsymbol{k};t,\boldsymbol{z}) = ex_{Ui}(\boldsymbol{k};t,\boldsymbol{x})U_{ij}^2(\boldsymbol{k};t,\boldsymbol{z}) \quad (4.6)$$

$$et_{Cij}(\boldsymbol{k};t,\boldsymbol{z}) = ex_{Cj}(\boldsymbol{k};t,\boldsymbol{y})C_{ij}^2(\boldsymbol{k};t,\boldsymbol{z}) \quad (4.7)$$

The similar relations define the second-degree buyers expected trades $\boldsymbol{et}_{ij}(t,z;\boldsymbol{l})$. The second-degree sellers expected trades of seller $\boldsymbol{et}_{ij}(\boldsymbol{k};t,z;\boldsymbol{l})$ and buyer $\boldsymbol{et}_{ij}(t,z;\boldsymbol{l})$ are additive functions and we use the procedure similar to (3.11-3.15) and (3.27-3.36) to define aggregate second-degree expected trades and their flows (Appendix B). We define collective sellers $Ex(\boldsymbol{k};t,z)$ (B.12-B14) and buyers $Ex(t,z;\boldsymbol{l})$ (B.15-B.17) expectations. Integral by economic domain (3.4, 3.5) introduce collective sellers $Ex(\boldsymbol{k};t)$ (B.20, B.210) and buyers $Ex(t;\boldsymbol{l})$ (B.22, B.23) expectations as function of time $t$ and type of expectation (4.3, 4.4). Below we use these expectations to model oscillations of price $p(2;t)$ (2.5; 2.6).

## 5. Price and volatility oscillations

In Appendix C we derive the system of equations on the second-degree transactions $\boldsymbol{B}(2;t,z)$ (3.11-3.15) and their flows $\boldsymbol{P}(2;t,z)$ (3.28-3.37) and on expected trades according to (Olkhov, 2019b). The system of equation (C.1-C.5) in economic domain (3.4, 3.5) and aggregated form of the equations (C.6-C.9) that describe collective transactions of the entire economy as functions of type of expectations $\boldsymbol{k}$ and time $t$ only are complex. We present the simplest consequence of the equations (C.5-C.9) to demonstrate possible origin of price fluctuations and present equations that describe price and volatility evolution.

In Olkhov (2019b) we show how simple equations on the first-degree transactions imply equations on the price $p(1,t)$ (1.2; 1.3). Here we derive the similar equations on the price $p(2,t)$ (2.5; 2.6) for $n=2$. To do that we neglect the action of the flows in (C.6, C.7) and take equations on second-degree trades as:

$$\frac{d}{dt}U(2,\boldsymbol{k};t) = F_U(\boldsymbol{k};t) \;\; ; \frac{d}{dt}C(2,\boldsymbol{k};t) = F_C(\boldsymbol{k};t) \tag{5.1}$$

Let's remind the form of the second-degree price $p(2,\boldsymbol{k};t)$ as

$$C(2,\boldsymbol{k};t) = p(2,\boldsymbol{k};t)U(2,\boldsymbol{k};t) \tag{5.2}$$

Equations (5.1, 5.2) on $U(2;\boldsymbol{k},t)$ and $C(2;\boldsymbol{k},t)$ allow present equations on $p(2;\boldsymbol{k},t)$ as:

$$U(2,\boldsymbol{k};t)\frac{d}{dt}p(2,\boldsymbol{k};t) + p(2,\boldsymbol{k};t)F_U(\boldsymbol{k};t) = F_C(\boldsymbol{k};t) \tag{5.3}$$

Equations (5.1, 5.3) show that dynamics of price $p(2,\boldsymbol{k};t)$ or similar equations (5.4) for price $p(2;t)$ and volume $U(2;t)$ (3.18, 3.19):

$$\frac{d}{dt}U(2;t) = F_U(t) \;\; ; \;\; U(2;t)\frac{d}{dt}p(2;t) + p(2;t)F_U(t) = F_C(t) \tag{5.4}$$

are determined by functions $F_U$, $F_C$. As we discussed above, $F_U$, $F_C$ can depend on expectations, expected transactions and their flows. Even in the simplest case if one neglects impact of transactions flows $P_U$, $P_C$ (A.7-A.10) and flows of expected transactions (B.28, B.29) and their equations (C.3, C.5) the equations (5.3, 5.4) on price $p(2,\boldsymbol{k};t)$ or price $p(2;t)$ depend on functions $F_U$, $F_C$ that model dynamics of collective the second-degree trades. Equations on *p(2,**k**;t)* or *p(2;t)* and similar equations on *p(1;t)* (Olkhov, 2019) describe evolution of price volatility $\sigma_p{}^2$ (2.10) determined by the first and the second-degree transactions.

Equations (5.1) allow describe fluctuations of the price *p(2,**k**;t)*. To do that assume that in linear approximation by perturbations functions $F_U$ and $F_C$ in (5.1) depend on perturbations of collective expected trades $Et_U(\boldsymbol{k};t)$ and $Et_C(\boldsymbol{k};t)$ (B.19) as (D.9). Let's take equations on expected trades $Et_U(\boldsymbol{k};t)$ and $Et_C(\boldsymbol{k};t)$ as (C.8) :

$$\frac{d}{dt}Et_U(\boldsymbol{k};t) = W_U(\boldsymbol{k};t) \ ; \quad \frac{d}{dt}Et_C(\boldsymbol{k};t) = W_C(\boldsymbol{k};t) \qquad (5.5)$$

and assume that in linear approximation by perturbations functions $W_U$ and $W_C$ in (5.5) depend on perturbations of the volume *U(2,**k**;t)* and the value *C(2,**k**;t)* as (D.10). Assumptions (5.1, 5.5) describe small oscillations of collective transactions *U(2,k;t)* and *C(2,k;t)* and hence describe fluctuations of price *p(2,**k**;t)* (5.2). We model fluctuations in Appendix D.

Relations (D.4) express the mean square price *p(2;t)* (3.19) disturbances though disturbances of the squares volumes *u(2,**k**;t)* (D.1) and the squares values *c(2,**k**;t)* (D.1). Relations (D.7, D.8) describe *p(2;t)* (3.19) dependence on partial price disturbances *π(2,**k**;t)* (D.6) and volume disturbances *u(2,**k**;t)* (D.1). Disturbances of *p(2;t)* (D.4, D.8) and disturbances of *p(1;t)* (D.26) describe price volatility $\sigma_p{}^2(t)$ (2.10) in the linear approximation by disturbances (D.27):

$$\sigma_p^2(t) = \sigma_{p0}^2\{1 + \sum_{k,l=1}^{K} \sigma_2[\mu_{2\boldsymbol{k}}c(2,\boldsymbol{k};t) - \lambda_{2\boldsymbol{k}}u(2,\boldsymbol{k};t)] - \sigma_1[\mu_{1\boldsymbol{k}}c(1,\boldsymbol{k};t) - \lambda_{1\boldsymbol{k}}u(1,\boldsymbol{k};t)]\}$$

We underline that above relations describe disturbances of price volatility $\sigma_p{}^2(t)$ (2.10) after the averaging price $p(t_i)$ by time interval *Δ*. In other words, if one average price instantaneous time-series $p(t_i)$, *i=1,2,…N(t)* by time interval *Δ* equals 1 day then during time terms equal weeks or months the price volatility $\sigma_p{}^2(t)$ (2.10) can follow above small perturbations.

We leave further development of these problems for future.

## 6. Conclusion

Price is a core notion of economics and finance and has many notions and definitions (Fetter, 1912). We don't advocate selection of any particular price definition but underline that the choice of price definition and the method for price aggregation determine result price properties. Instantaneous, momentary price have little usage due to irregular price oscillations. Investors, traders and researchers aggregate price during certain time interval *Δ* and use smooth averaged price function. The choice of interval $\Delta$ and the choice of averaging procedure play most important role for successive investment and stable portfolio returns, for adequate macro financial modeling and sustainable financial policy.

We propose that most common method for price averaging that is based on price $p$ frequencies $n_p$ (1.1) is not the only one and not the best one for investors and economic policy management. Investment decisions and investors portfolio strategies should follow valuable and significant market trades, those have most market impact.

We propose unit market price measure $\eta(p;t)$ (2.7-2.9) that for all $n=1,2,...$ provide $n$-$th$ price statistical moments $p(n;t)$ with respect to measure $\eta(p;t)$ in form (2.5)

$$C(n;t) = p(n;t)U(n;t)$$

that reproduce relations between the value $C$, the volume $U$ and the price $p$ of the single market trade (2.1). These relations allow present the price volatility $\sigma_p^2(t)$ (2.10)

$$\sigma_p^2(t) = p(2;t) - p^2(1,t) = \frac{C(2;t)}{U(2;t)} - \frac{C^2(1;t)}{U^2(1;t)}$$

as function of first and second degree values and volumes of market trades. Forecasting of market price volatility $\sigma_p^2(t)$ (2.10) with respect to unit measure $\eta(p;t)$ requires description of second-degree macro trades. It seems reasonably enough that price volatility should depend on evolution of market trades and macro finance. We start modeling dynamics of the second-degree market trades. It is important that aggregation of market trades and transition to continuous media approximation within averaging of market trades during the time interval $\Delta$ (see App. A-D) derive the aggregated price $p(2;t,z)$ (3.16) absolutely similar to definition to $p(2;t)$ (2.5). That is additional confirmation of correctness of our proposition to use the unit price measure $\eta(p;t)$ (2.7-2.9) as the tool to derive mean price, price volatility and all $n$-$th$ price statistical moments. We take VWAP as initial point of our approach and introduce mean $n$-$th$ degree price $p(n;t)$ (2.4-2.6).

Relations (2.4-2.9) show that price measure $\eta(p;t)$ depends on sum of the $n$-$th$ degree of the value $C(n;t)$ and the volume $U(n;t)$ (2.4) aggregated during interval $\Delta(t)$ (1.4). As we show above, description of $C(2;t)$ and $U(2;t)$ requires development of the second-order theory that models dynamics of second-degree value $C(2;t)$ and volume $U(2;t)$. Accordingly description of $n$-$th$ degree value $C(n;t)$ and volume $U(n;t)$ requires development of the $n$-$th$ order theories for all n=1,2,… This is sufficiently tough problem that unlikely be solved ever. That make exact forecasting of the unit price measure $\eta(p;t)$ almost unreal. Fortunately or unfortunately that makes hopes of exact forecast of the market price probability unattainable. However the development of the second-order economic theory that model second-degree trades may help for reasonable forecasting of the price volatility $\sigma_p^2(t)$ (2.10).

However investors, traders and researchers may rely on their own price expectations, price hopes and anticipations and use any other possible price averaging procedure, different from the unit measure $\eta(p;t)$ (2.7-2.9). Economics is a social science and in our view no firm rules and regulations for price averaging methods may exist. The distinctions between various price definitions and price averaging methods have significant effect on economic and financial modeling, market forecasting and investment strategies, portfolio optimization and hedging models. Only comparisons between portfolio returns results obtained via investment strategies based on different price averaging measures and procedures can convince investors in the advantages of the unit measure $\eta(p;t)$.

We hope that further development of proposed unit price measure $\eta(p;t)$ and the second-order economic theory can bring useful financial solutions, improve investment strategies and sustainability of portfolio returns.

## Appendix A. Transactions and their flows

Integral of the second-degree trade $\boldsymbol{B}(2;t,z)$ (3.11-3.15) by $d\boldsymbol{y}$ over economic domain (3.2) determines all sell-trades $\boldsymbol{B}_s(2;t,\boldsymbol{x})$ from point $\boldsymbol{x}$ :

$$\boldsymbol{B}_s(2;t,\boldsymbol{x}) = \int d\boldsymbol{y}\ \boldsymbol{B}(2;t,\boldsymbol{x},\boldsymbol{y}) = \big(U_s(2;t,\boldsymbol{x});C_s(2;t,\boldsymbol{x})\big) \tag{A.1}$$

$$U_s(2;t,\boldsymbol{x}) = \int d\boldsymbol{y}\ U(2;t,\boldsymbol{x},\boldsymbol{y})\ \ ;\ \ C_s(2;t,\boldsymbol{x}) = \int d\boldsymbol{y}\ C(2;t,\boldsymbol{x},\boldsymbol{y}) \tag{A.2}$$

Integral $\boldsymbol{B}(2;t,z)$ by $d\boldsymbol{x}$ over economic domain (3.2) determines all buy-trades $\boldsymbol{B}_b(2;t,\boldsymbol{y})$ at $\boldsymbol{y}$

$$\boldsymbol{B}_b(2;t,\boldsymbol{y}) = \int d\boldsymbol{x}\ \boldsymbol{B}(2;t,\boldsymbol{x},\boldsymbol{y}) = \big(U_b(2;t,\boldsymbol{y});C_b(2;t,\boldsymbol{y})\big) \tag{A.3}$$

$$U_b(2;t,\boldsymbol{y}) = \int d\boldsymbol{x}\ U(2;t,\boldsymbol{x},\boldsymbol{y})\ \ ;\ \ C_b(2;t,\boldsymbol{y}) = \int d\boldsymbol{x}\ C(2;t,\boldsymbol{x},\boldsymbol{y}) \tag{A.4}$$

Integral of the second-degree trade $B(2;t,z)$ by $dz=d\boldsymbol{x}d\boldsymbol{y}$ over economic domain (3.2) define all macro second-degree trades $\boldsymbol{B}(2;t)$ as function of time $t$ only:

$$\boldsymbol{B}(2;t) = \int d\boldsymbol{z}\ \boldsymbol{B}(2;t,\boldsymbol{z}) = \big(U(2;t);C(2;t)\big) \tag{A.5}$$

$$U(2;t) = \int d\boldsymbol{z}\ U(2;t,\boldsymbol{z})\ \ ;\ \ C(2;t) = \int d\boldsymbol{z}\ C(2;t,\boldsymbol{z}) \tag{A.6}$$

Relations (A.1-A.6) introduce flows and velocities for corresponding second-degree trades. Due to (3.31-3.32) obtain:

$$\boldsymbol{P}_{Ux}(2;t,\boldsymbol{x}) = \int d\boldsymbol{y}\,\boldsymbol{P}_{Ux}(2;t,\boldsymbol{x},\boldsymbol{y}) = \textstyle\sum_{i\in dV(\boldsymbol{x});j\in dV(\boldsymbol{y})\,\Delta}\int d\boldsymbol{y}\,U_{ij}^2(t,\boldsymbol{x},\boldsymbol{y})\boldsymbol{v}_i(t,\boldsymbol{x}) \tag{A.7}$$

$$\boldsymbol{P}_{Ux}(2;t,\boldsymbol{x}) = \int d\boldsymbol{y}\ U(2;t,\boldsymbol{z})\ \boldsymbol{v}_{Ux}(2;t,\boldsymbol{z}) = U_s(2;t,\boldsymbol{x})\boldsymbol{v}_{Usx}(2;t,\boldsymbol{x}) \tag{A.8}$$

$$\boldsymbol{P}_{Uy}(2;t,\boldsymbol{x}) = \int d\boldsymbol{y}\,\boldsymbol{P}_{Uy}(2;t,\boldsymbol{x},\boldsymbol{y}) = \textstyle\sum_{i\in dV(\boldsymbol{x});j\in dV(\boldsymbol{y})\,\Delta}\int d\boldsymbol{y}\,U_{ij}^2(t,\boldsymbol{x},\boldsymbol{y})\boldsymbol{v}_j(t,\boldsymbol{y}) \tag{A.9}$$

$$\boldsymbol{P}_{Uy}(2;t,\boldsymbol{x}) = \int d\boldsymbol{y}\ U(2;t,\boldsymbol{z})\ \boldsymbol{v}_{Uy}(2;t,\boldsymbol{z}) = U_s(2;t,\boldsymbol{x})\boldsymbol{v}_{Uby}(2;t,\boldsymbol{x}) \tag{A.10}$$

Relations (A.7, A.8) introduce aggregate flow of sales $\boldsymbol{P}_{Ux}(2;t,\boldsymbol{x})$ of the second-degree volumes $U_s(2;t,\boldsymbol{x})$ and collective velocity $\boldsymbol{v}_{Usx}(2;t,\boldsymbol{x})$ of sellers at point $\boldsymbol{x}$ along axis X to all buyers in economy. Relations (A.9, A.10) introduce aggregate flow of sales $\boldsymbol{P}_{Uy}(2;t,\boldsymbol{x})$ of second-degree volumes $U_s(2;t,\boldsymbol{x})$ and collective velocity $\boldsymbol{v}_{Usy}(2;t,\boldsymbol{x})$ of all buyers in economy along axis Y from collective sellers at point $\boldsymbol{x}$. Corresponding relations for the second-degree values flows $\boldsymbol{P}_{Csx}(2;t,\boldsymbol{x})$ and $\boldsymbol{P}_{Csy}(2;t,\boldsymbol{x})$ take form:

$$\boldsymbol{P}_{Csx}(2;t,\boldsymbol{x}) = \int d\boldsymbol{y}\,\boldsymbol{P}_{Cx}(2;t,\boldsymbol{x},\boldsymbol{y}) = \textstyle\sum_{i\in dV(\boldsymbol{x});j\in dV(\boldsymbol{y})\,\Delta}\int d\boldsymbol{y}\,C_{ij}^2(t,\boldsymbol{x},\boldsymbol{y})\boldsymbol{v}_i(t,\boldsymbol{x}) \tag{A.11}$$

$$\boldsymbol{P}_{Csx}(2;t,\boldsymbol{x}) = C_s(2;t,\boldsymbol{x})\boldsymbol{v}_{Csx}(2;t,\boldsymbol{x}) \tag{A.12}$$

$$\boldsymbol{P}_{Csy}(2;t,\boldsymbol{x}) = \int d\boldsymbol{y}\,\boldsymbol{P}_{Cy}(2;t,\boldsymbol{x},\boldsymbol{y}) = \textstyle\sum_{i\in dV(\boldsymbol{x});j\in dV(\boldsymbol{y})\,\Delta}\int d\boldsymbol{y}\,C_{ij}^2(t,\boldsymbol{x},\boldsymbol{y})\boldsymbol{v}_j(t,\boldsymbol{y}) \tag{A.13}$$

$$\boldsymbol{P}_{Csy}(2;t,\boldsymbol{x}) = C_s(2;t,\boldsymbol{x})\boldsymbol{v}_{Csy}(2;t,\boldsymbol{x}) \tag{A.14}$$

The similar relations define the buyer's flows. To derive buyer's the second-degree volume and the value flows $\boldsymbol{P}_{Ubx}(2;t,\boldsymbol{y})$ and $\boldsymbol{P}_{Uby}(2;t,\boldsymbol{y})$, $\boldsymbol{P}_{Cbx}(2;t,\boldsymbol{y})$ and $\boldsymbol{P}_{Cby}(2;t,\boldsymbol{y})$ one should take integral for (3.32-3.35) by $d\boldsymbol{x}$ over economic domain (3.4, 3.5). For brevity we omit it here.

Integrals by $d\boldsymbol{x}d\boldsymbol{y}$ for (3.31-3.35) over entire economy – over economic domain (3.4,3.5) introduce macro flows of second-degree transactions as functions of time $t$ only:

$$\boldsymbol{P}_{Ux}(2;t) = \int d\boldsymbol{z}\, \boldsymbol{P}_{Ux}(2;t,\boldsymbol{z}) = \sum_{i\in dV(\boldsymbol{x});j\in dV(\boldsymbol{y})}{}_{\Delta} \int d\boldsymbol{x}d\boldsymbol{y}\, U_{ij}^{2}(t,\boldsymbol{x},\boldsymbol{y})\boldsymbol{v}_i(t,\boldsymbol{x}) \quad (A.15)$$

$$\boldsymbol{P}_{Ux}(2;t) = U(2;t)\boldsymbol{v}_{Ux}(2;t) \quad (A.16)$$

$$\boldsymbol{P}_{Uy}(2;t) = \int d\boldsymbol{z}\, \boldsymbol{P}_{Uy}(2;t,\boldsymbol{z}) = \sum_{i\in dV(\boldsymbol{x});j\in dV(\boldsymbol{y})}{}_{\Delta} \int d\boldsymbol{x}d\boldsymbol{y}\, U_{ij}^{2}(t,\boldsymbol{x},\boldsymbol{y})\boldsymbol{v}_j(t,\boldsymbol{y}) \quad (A.17)$$

$$\boldsymbol{P}_{Uy}(2;t) = U(2;t)\boldsymbol{v}_{Uy}(2;t) \quad (A.18)$$

$$\boldsymbol{P}_{Cx}(2;t) = \int d\boldsymbol{z}\, \boldsymbol{P}_{Cx}(2;t,\boldsymbol{z}) = \sum_{i\in dV(\boldsymbol{x});j\in dV(\boldsymbol{y})}{}_{\Delta} \int d\boldsymbol{x}d\boldsymbol{y}\, C_{ij}^{2}(t,\boldsymbol{x},\boldsymbol{y})\boldsymbol{v}_i(t,\boldsymbol{x}) \quad (A.19)$$

$$\boldsymbol{P}_{Cx}(2;t) = C(2;t)\boldsymbol{v}_{Cx}(2;t) \quad (A.20)$$

$$\boldsymbol{P}_{Cy}(2;t) = \int d\boldsymbol{z}\, \boldsymbol{P}_{Cy}(2;t,\boldsymbol{z}) = \sum_{i\in dV(\boldsymbol{x});j\in dV(\boldsymbol{y})}{}_{\Delta} \int d\boldsymbol{x}d\boldsymbol{y}\, C_{ij}^{2}(t,\boldsymbol{x},\boldsymbol{y})\boldsymbol{v}_j(t,\boldsymbol{y}) \quad (A.21)$$

$$\boldsymbol{P}_{Cy}(2;t) = C(2;t)\boldsymbol{v}_{Cy}(2;t) \quad (A.22)$$

Relations (A.15-A.22) introduce new economic concepts – collective flows of all agents of the economy involved into the second-degree trades in economic domain (3.4, 3.5). Collective flows of the second-degree volumes ($\boldsymbol{P}_{Ux}(2;t)$; $\boldsymbol{P}_{Uy}(2;t)$), and flows of the second-degree values ($\boldsymbol{P}_{Cx}(2;t)$; $\boldsymbol{P}_{Cy}(2;t)$) define corresponding collective velocities of the volume ($\boldsymbol{v}_{Ux}$, $\boldsymbol{v}_{Uy}$) and the value ($\boldsymbol{v}_{Cx}$, $\boldsymbol{v}_{Cy}$). Velocities $\boldsymbol{v}_{Ux}$ define collective motion of all sellers in economy weighted by volumes of trades during interval $\Delta$. Velocities $\boldsymbol{v}_{Uy}$ define collective motion of all buyers in economy weighted by volumes of trades during interval $\Delta$. Velocities ($\boldsymbol{v}_{Cx}$, $\boldsymbol{v}_{Cy}$) define collective motion of all sellers and all buyers weighted by value of all trades during interval $\Delta$. These velocities describe collective motion of the entire economy in the economic domain (3.4, 3.5) reduced by its borders at 0 and 1 by all axes. Hence these collective velocities can't have the constant direction and must oscillate. Collective velocities of the volume ($\boldsymbol{v}_{Ux}$, $\boldsymbol{v}_{Uy}$) and the value ($\boldsymbol{v}_{Cx}$, $\boldsymbol{v}_{Cy}$) occasionally change their direction and describe collective motion of sellers and buyers from the secure area of economic domain to the risky one and back. We suggest that such fluctuations of the second-degree trades accompanied with rise and fall of market activity, production functions, demand and supply, price variations and etc., establish ground for economic fluctuations of business cycles. We describe motion of collective trades in the economic domain (3.4, 3.5) as hidden properties of the business cycles (Olkhov, 2017b, 2019a).

## Appendix B. Collective expectations

To define collective expectations we introduce new issue: collective expected trades. To do that we use (4.1) to define collective second-degree trades $U(2,\boldsymbol{k};t,z)$, $C(2,\boldsymbol{k};t,z)$ approved by sellers expectations of type $\boldsymbol{k}=(k,l)$ as:

$$U(2,\boldsymbol{k};t,\boldsymbol{z}) = \sum_{i\in dV(\boldsymbol{x});j\in dV(\boldsymbol{y})}{}_{\Delta} U_{ij}^{2}(\boldsymbol{k};t,\boldsymbol{z}) \quad (B.1)$$

$$C(2,\boldsymbol{k};t,\boldsymbol{z}) = \sum_{i\in dV(\boldsymbol{x});j\in dV(\boldsymbol{y})}{}_{\Delta} C_{ij}^{2}(\boldsymbol{k};t,\boldsymbol{z}) \quad (B.2)$$

Definitions of collective flows $\boldsymbol{P}_U(2,\boldsymbol{k};t,z)$ and $P_C(2,\boldsymbol{k};t,z)$ of trades with the volume $U(2,\boldsymbol{k};t,z)$ and the value $C(2,\boldsymbol{k};t,z)$ (B.1,B.2) are similar to (A.7-A.10) and we omit it here for brevity. Trades $U(2,\boldsymbol{k};t,z)$, $C(2,\boldsymbol{k};t,z)$ (B.1, B.2) obey obvious relations (3.13-3.15):

$$U(2;t,\boldsymbol{z}) = \sum_{k,l=1}^{K} U(2,\boldsymbol{k};t,\boldsymbol{z}) \;\; ; \;\; C(2;t,\boldsymbol{z}) = \sum_{k,l=1}^{K} C(2,\boldsymbol{k};t,\boldsymbol{z}) \tag{B.3}$$

Collective buyers second-degree trades $U(2;t,z;\boldsymbol{l})$, $C(2;t,z;\boldsymbol{l})$ approved by buyers expectations take the similar form:

$$U(2;t,\boldsymbol{z};\boldsymbol{l}) = \sum_{i\in dV(\boldsymbol{x});j\in dV(\boldsymbol{y})\,\Delta} U_{ij}^2(t,\boldsymbol{z};\boldsymbol{l}) \tag{B.4}$$

$$C(2;t,\boldsymbol{z};\boldsymbol{l}) = \sum_{i\in dV(\boldsymbol{x});j\in dV(\boldsymbol{y})\,\Delta} C_{ij}^2(t,\boldsymbol{z};\boldsymbol{l}) \tag{B.5}$$

Buyer's trades obey same relations (B.3). Similar to (3.11-3.15) we use (4.5-4.7) and define sellers collective expected trades $\boldsymbol{Et}(\boldsymbol{k};t,z)$ as:

$$\boldsymbol{Et}(k;t,\boldsymbol{z}) = \big(Et_U(\boldsymbol{k};t,\boldsymbol{z}); Et_C(\boldsymbol{k};t,\boldsymbol{z})\big) \tag{B.6}$$

$$Et_U(\boldsymbol{k};t,\boldsymbol{z}) = \sum_{i\in dV(\boldsymbol{x});j\in dV(\boldsymbol{y});\,\Delta} ex_{Ui}(\boldsymbol{k};t,\boldsymbol{x})U_{ij}^2(\boldsymbol{k};t,\boldsymbol{z}) \tag{B.7}$$

$$Et_C(\boldsymbol{k};t,\boldsymbol{z}) = \sum_{i\in dV(\boldsymbol{x});j\in dV(\boldsymbol{y});\,\Delta} ex_{Ci}(\boldsymbol{k};t,\boldsymbol{x})C_{ij}^2(\boldsymbol{k};t,\boldsymbol{z}) \tag{B.8}$$

Buyer's collective second-degree expected trades $\boldsymbol{Et}(t,z;\boldsymbol{l})$,

$$\boldsymbol{Et}(t,\boldsymbol{z};\boldsymbol{l}) = \big(Et_U(t,\boldsymbol{z};\boldsymbol{l}); Et_C(t,\boldsymbol{z};\boldsymbol{l})\big) \tag{B.9}$$

$$Et_U(t,\boldsymbol{z};\boldsymbol{l}) = \sum_{i\in dV(\boldsymbol{x});j\in dV(\boldsymbol{y});\,\Delta} ex_{Uj}(t,\boldsymbol{y};\boldsymbol{l})U_{ij}^2(t,\boldsymbol{z};\boldsymbol{l}) \tag{B.10}$$

$$Et_C(t,\boldsymbol{z};\boldsymbol{l}) = \sum_{i\in dV(\boldsymbol{x});j\in dV(\boldsymbol{y});\,\Delta} ex_{Cj}(t,\boldsymbol{y};\boldsymbol{l})C_{ij}^2(t,\boldsymbol{z};\boldsymbol{l}) \tag{B.11}$$

Relations (B.6-B.8) and (B.1,B.2) introduce collective sellers expectations of the second-degree $\boldsymbol{Ex}(\boldsymbol{k};t,z)$, as:

$$Ex(\boldsymbol{k};t,\boldsymbol{z}) = \big(Ex_U(\boldsymbol{k};t,\boldsymbol{z}), Ex_C(\boldsymbol{k};t,\boldsymbol{z})\big) \tag{B.12}$$

$$Ex_U(\boldsymbol{k};t,\boldsymbol{z})U(2,\boldsymbol{k};t,\boldsymbol{z}) = Et_U(\boldsymbol{k};t,\boldsymbol{z}) \tag{B.13}$$

$$Ex_C(\boldsymbol{k};t,\boldsymbol{z})C(2,\boldsymbol{k};t,\boldsymbol{z}) = Et_C(\boldsymbol{k};t,\boldsymbol{z}) \tag{B.14}$$

Collective buyers expectations take similar form:

$$Ex(t,\boldsymbol{z};\boldsymbol{l}) = \big(Ex_U(t,\boldsymbol{z};\boldsymbol{l}), Ex_C(t,\boldsymbol{z};\boldsymbol{l})\big) \tag{B.15}$$

$$Ex_U(t,\boldsymbol{z};\boldsymbol{l})U(2;t,\boldsymbol{z};\boldsymbol{l}) = Et_U(t,\boldsymbol{z};\boldsymbol{l}) \tag{B.16}$$

$$Ex_C(t,\boldsymbol{z};\boldsymbol{l})C(2;t,\boldsymbol{z};\boldsymbol{l}) = Et_C(t,\boldsymbol{z};\boldsymbol{l}) \tag{B.17}$$

Let's take integrals over economic domain (3.4,3.5) of $U(2,\boldsymbol{k};t,z)$, $C(2,\boldsymbol{k};t,z)$ (B.1, B.2):

$$U(2,\boldsymbol{k};t) = \int d\boldsymbol{z}\; U(2,\boldsymbol{k};t,\boldsymbol{z})\,; \;\; C(2,\boldsymbol{k};t) = \int d\boldsymbol{z}\; C(2,\boldsymbol{k};t,\boldsymbol{z}) \tag{B.18}$$

Relations (B.18) define collective sellers second-degree trades $U(2,\boldsymbol{k};t)$, $C(2,\boldsymbol{k};t)$ performed under expectations of type $\boldsymbol{k}=(k,l)$. Then integral by economic domain (3.4, 3.5) of expected trades (B.13, B.14) define collective sellers expected trades $Et_U(\boldsymbol{k};t)$, $Et_C(\boldsymbol{k};t)$ of the entire economy

$$Et_U(\boldsymbol{k};t) = \int d\boldsymbol{z}\ Et_U(\boldsymbol{k};t,\boldsymbol{z})\ ;\ \ Et_C(\boldsymbol{k};t) = \int d\boldsymbol{z}\ Et_C(\boldsymbol{k};t,\boldsymbol{z}) \tag{B.19}$$

(B.18, B.19) define collective sellers expectations $Ex_U(\boldsymbol{k};t)$, $Ex_C(\boldsymbol{k};t)$ of the entire economy

$$Ex_U(\boldsymbol{k};t)U(2,\boldsymbol{k};t) = Et_U(\boldsymbol{k};t) \tag{B.20}$$

$$Ex_C(\boldsymbol{k};t)C(2,\boldsymbol{k};t) = Et_C(\boldsymbol{k};t) \tag{B.21}$$

The similar relations define collective buyers expected trades $Et_U(t;\boldsymbol{l})$, $Et_C(t;\boldsymbol{l})$ and expectations $Ex_U(t;\boldsymbol{l})$, $Ex_C(t;\boldsymbol{l})$ of the entire economy:

$$Ex_U(t;\boldsymbol{l})U(2;t;\boldsymbol{l}) = Et_U(t;\boldsymbol{l}) \tag{B.22}$$

$$Ex_C(t;\boldsymbol{l})C(2;t;\boldsymbol{l}) = Et_C(t;\boldsymbol{l}) \tag{B.23}$$

Relations (B.20) introduce collective seller expected trades $Et_{Us}(t)$, $Et_{Cs}(t)$:

$$Et_{Us}(t) = \sum_{k,l=1}^{K} Et_U(\boldsymbol{k};t)\ \ ;\ \ Et_{Cs}(t) = \sum_{k,l=1}^{K} Et_C(\boldsymbol{k};t) \tag{B.24}$$

Relations (B.24) define collective sellers expectations $Ex_{Us}(t)$, $Ex_{Cs}(t)$:

$$Ex_{Us}(t)U(2;t)\ = Et_{Us}(t)\ \ ;\ \ Ex_{Cs}(t)C(2;t)\ = Et_{Cs}(t) \tag{B.25}$$

The similar relations define collective buyers expectations of the entire economy. We omit them for brevity. It is clear that collective sellers expectations those approve squares of the volume $U^2$ and squares of the value $C^2$ of transactions may be different. Collective seller's expectations may be different from buyer's expectations. Such diversity uncovers hidden complexity of the mutual interactions between trades and expectations. We present above set of long definitions to outline multiplicity of different states of collective trades and expectations. One easy defines collective flows of sellers and buyers expected trades similar to (Olkhov, 2019b). For brevity we present only definition of sellers volume flow of expected trades $\boldsymbol{Pe}_{Ux}$, $\boldsymbol{Pe}_{Uy}$ in (B.26-B2.9):

$$\boldsymbol{pe}_{Uxij}(\boldsymbol{k};t,\boldsymbol{z}) = et_{Uij}(\boldsymbol{k};t,\boldsymbol{z})\boldsymbol{v}_i(t,\boldsymbol{x}) = ex_i(\boldsymbol{k};t,\boldsymbol{x})U_{ij}^2(\boldsymbol{k};t,\boldsymbol{z})\boldsymbol{v}_i(t,\boldsymbol{x}) \tag{B.26}$$

$$\boldsymbol{pe}_{Uyij}(\boldsymbol{k};t,\boldsymbol{z}) = et_{Uij}(\boldsymbol{k};t,\boldsymbol{z})\boldsymbol{v}_j(t,\boldsymbol{y}) = ex_i(\boldsymbol{k};t,\boldsymbol{x})U_{ij}^2(\boldsymbol{k};t,\boldsymbol{z})\boldsymbol{v}_j(t,\boldsymbol{y}) \tag{B.27}$$

$$\boldsymbol{Pe}_{Ux}(\boldsymbol{k};t,\boldsymbol{z}) = \sum_{i\in dV(x);j\in dV(y)} {}_{\Delta}et_{Uij}(\boldsymbol{k};t,\boldsymbol{z})\boldsymbol{v}_i(t,\boldsymbol{x}) = Et_U(\boldsymbol{k};t,\boldsymbol{z})\boldsymbol{ve}_{Ux}(\boldsymbol{k};t,\boldsymbol{z}) \tag{B.28}$$

$$\boldsymbol{Pe}_{Uy}(\boldsymbol{k};t,\boldsymbol{z}) = \sum_{i\in dV(x);j\in dV(y)} {}_{\Delta}et_{Uij}(\boldsymbol{k};t,\boldsymbol{z})\boldsymbol{v}_j(t,\boldsymbol{y}) = Et_U(\boldsymbol{k};t,\boldsymbol{z})\boldsymbol{ve}_{Uy}(\boldsymbol{k};t,\boldsymbol{z}) \tag{B.29}$$

For brevity we don't' define full set of flows and refer (Olkhov, 2019b) for further detail.

## Appendix C. Equations of the second-degree trades

For brevity we present here only equations on second-degree trades ***B****(2;t,z)* (3.11-3.15) and refer for details (Olkhov, 2018-2019b). Equations on the second-degree volume *U(2,**k**;t,z)* and the value *C(2,**k**;t,z)* (B.1,B.2) take form:

$$\frac{\partial}{\partial t}U(2,\boldsymbol{k};t,\boldsymbol{z}) + \nabla\cdot\big(U(2,\boldsymbol{k};t,\boldsymbol{z})\ \boldsymbol{v}_U(2,\boldsymbol{k};t,\boldsymbol{z})\big) = F_U(\boldsymbol{k};t,\boldsymbol{z}) \tag{C.1}$$

$$\frac{\partial}{\partial t}C(2,\boldsymbol{k};t,\boldsymbol{z}) + \nabla\cdot\big(C(2,\boldsymbol{k};t,\boldsymbol{z})\,\boldsymbol{v}_C(2,\boldsymbol{k};t,\boldsymbol{z})\big) = F_C(\boldsymbol{k};t,\boldsymbol{z}) \qquad \text{(C.2)}$$

Equations on flows $\boldsymbol{P}_{Ux}(2;t,z)$ (3.32) of second-degree volume $U(2;t,z)$ take form:

$$\frac{\partial}{\partial t}\boldsymbol{P}_{Ux}(\boldsymbol{k};t,\boldsymbol{z}) + \nabla\cdot\big(\boldsymbol{P}_{Ux}(\boldsymbol{k};t,\boldsymbol{z})\,\boldsymbol{v}_{Ux}(\boldsymbol{k};t,\boldsymbol{z})\big) = \boldsymbol{G}_{Ux}(\boldsymbol{k};t,\boldsymbol{z}) \qquad \text{(C.3)}$$

Equations on flows (3.33-3.35) have similar form. Similar equations valid for expected trades $Ex_U(\boldsymbol{k};t,z)$ (B.6-B.8) and their flows $\boldsymbol{Pe}_{Ux}(\boldsymbol{k};t,\boldsymbol{z})$:

$$\frac{\partial}{\partial t}Et_U(\boldsymbol{k};t,\boldsymbol{z}) + \nabla\cdot\big(Et_U(\boldsymbol{k};t,\boldsymbol{z})\,\boldsymbol{ve}_{Ux}(\boldsymbol{k};t,\boldsymbol{z})\big) = W_U(\boldsymbol{k};t,\boldsymbol{z}) \qquad \text{(C.4)}$$

$$\frac{\partial}{\partial t}\boldsymbol{Pe}_{Ux}(\boldsymbol{k};t,\boldsymbol{z}) + \nabla\cdot\big(\boldsymbol{Pe}_{Ux}(\boldsymbol{k};t,\boldsymbol{z})\,\boldsymbol{ve}_{Ux}(\boldsymbol{k};t,\boldsymbol{z})\big) = \boldsymbol{R}_{Ux}(\boldsymbol{k};t,\boldsymbol{z}) \qquad \text{(C.5)}$$

Integrals of divergence by $dz$ over domain (3.4,3.5) equal zero as no economic agents and no fluxes exist outside (3.4,3.5). Hence integrals of (C.1-C.5) by $dz$ over domain (3.4,3.5) give:

$$\int d\boldsymbol{z}\left[\frac{\partial}{\partial t}U(2,\boldsymbol{k};t,\boldsymbol{z}) + \nabla\cdot\big(U(2,\boldsymbol{k};t,\boldsymbol{z})\,\boldsymbol{v}_U(2,\boldsymbol{k};t,\boldsymbol{z})\big)\right] = \frac{d}{dt}U(2,\boldsymbol{k};t) = F_U(\boldsymbol{k};t) = \int d\boldsymbol{z}\,F_U(t,\boldsymbol{z}) \qquad \text{(C.6)}$$

$$\frac{d}{dt}C(2,\boldsymbol{k};t) = F_C(\boldsymbol{k};t) \quad ; \quad \frac{d}{dt}\boldsymbol{P}_{Ux}(\boldsymbol{k};t) = \boldsymbol{G}_{Ux}(\boldsymbol{k};t) \qquad \text{(C.7)}$$

$$\frac{d}{dt}Et_U(\boldsymbol{k};t) = W_U(\boldsymbol{k};t) \quad ; \quad \frac{d}{dt}\boldsymbol{Pe}_{Ux}(\boldsymbol{k};t) = \boldsymbol{R}_{Ux}(\boldsymbol{k};t) \qquad \text{(C.8)}$$

(C.6-C.8) describe equations on collective second-degree trades, expectations and their flows as functions of time $t$ and type of expectations $\boldsymbol{k}$. Formal simplicity of (C.6-C.8) hides complexity of the functions $F_U(t)$, $F_U(t)$, $\boldsymbol{G}_{Ux}$, $\boldsymbol{W}_{Ux}$, $\boldsymbol{R}_{Ux}$, and etc. These functions describe impact of expectations of the second-degree trades and their flows and impact of trades on the expected trades and their flows. These functions model the economic and financial processes, social expectations, technology forecast, and flows of these parameters those impact the second-degree trades $\boldsymbol{B}(2,t)$ (3.17) and flows $\boldsymbol{P}(2,t)$ and back action of trades on expected trades and their flows. Integrals in right side of (C.6-C.8) can describe integrals by flows and their products. Such factors are completely new for macroeconomic theory and were never taken into account before.

## Appendix D. Price fluctuations

Let's follow (Olkhov, 2019b) and neglect the impact of trades and expected trades flows. Let's study small perturbations of $U(2,\boldsymbol{k};t)$, $C(2,\boldsymbol{k};t)$ and $Ex_U(\boldsymbol{k};t)$ and $Ex_C(\boldsymbol{k};t)$ as:

$$U(2,\boldsymbol{k};t) = U_{\boldsymbol{k}}^2\big(1 + u(2,\boldsymbol{k};t)\big)\ ;\ C(2,\boldsymbol{k};t) = C_{\boldsymbol{k}}^2\big(1 + c(2,\boldsymbol{k};t)\big) \qquad \text{(D.1)}$$

$$Et_U(\boldsymbol{k};t) = Et_{U\boldsymbol{k}}\big(1 + et_u(\boldsymbol{k};t)\big);\ Et_C(\boldsymbol{k};t) = Et_{C\boldsymbol{k}}\big(1 + et_c(\boldsymbol{k};t)\big) \qquad \text{(D.2)}$$

Relations (D.1, D.2) describe small dimensionless disturbances $u, c, et_u\ et_c$. Relations (3.19, 5.2, D.1) present the second-degree volume $U(2;t)$, the value $C(2;t)$ and price $p(2,t)$ (3.19) as:

$$U(2;t) = \sum_{k,l=1}^{K} U(2,\boldsymbol{k};t)\ ;\ C(2;t) = \sum_{k,l=1}^{K} C(2,\boldsymbol{k};t)\ ;\ C(2;t) = p(2;t)U(2;t) \qquad \text{(D.3)}$$

$$p(2;t) = \frac{C(2;t)}{U(2;t)} = \frac{\sum_{k,l} C_{\boldsymbol{k}}^2 \big(1 + c(2, \boldsymbol{k}; t)\big)}{\sum_{k,l} U_{\boldsymbol{k}}^2 \big(1 + u(2, \boldsymbol{k}; t)\big)}$$

In linear approximation by the disturbances $u(2,\boldsymbol{k};t)$ and $c(2,\boldsymbol{k};t)$, the price $p(2;t)$ takes form:

$$p(2;t) = p_{20}[1 + \pi(2;t)] = p_{20}\big[1 + \textstyle\sum_{k,l=1}^{K} (\mu_{2\boldsymbol{k}} c(2, \boldsymbol{k}; t) - \lambda_{2\boldsymbol{k}} u(2, \boldsymbol{k}; t))\big] \tag{D.4}$$

$$U_{20} = \sum\nolimits_{k,l} U_{\boldsymbol{k}}^2 \ ; \ C_{20} = \sum\nolimits_{k,l} C_{\boldsymbol{k}}^2 \ ; \quad C_{20} = p_{20} U_{20}$$

$$\lambda_{2\boldsymbol{k}} = \frac{U_{\boldsymbol{k}}^2}{U_{20}} \ ; \quad \mu_{2\boldsymbol{k}} = \frac{C_{\boldsymbol{k}}^2}{C_{20}} \ ; \ \sum \lambda_{2\boldsymbol{k}} = \sum \mu_{2\boldsymbol{k}} = 1$$

We use index *2* to underline that $p_{20}$ and other constants describe properties of the price $p(2;t)$. Let's take into account (D.1) and present price $p(2,\boldsymbol{k},t)$ (5.2) disturbances as

$$C_{\boldsymbol{k}}^2[1 + c(2, \boldsymbol{k}; t)] = p_{2k}[1 + \pi(2, \boldsymbol{k}; t)] U_{\boldsymbol{k}}^2 [1 + u(2, \boldsymbol{k}; t)] \tag{D.5}$$

Then in the linear approximation by disturbances obtain:

$$C_{\boldsymbol{k}}^2 = p_{2k} U_{\boldsymbol{k}}^2 \ ; \ \pi(2; \boldsymbol{k}; t) = c(2, \boldsymbol{k}; t) - u(2, \boldsymbol{k}; t) \tag{D.6}$$

Now substitute (D.6) into (D.4) and obtain dependence of price disturbances $\pi(2;t)$ on the volume disturbances $u(2,\boldsymbol{k};t)$:

$$\pi(2;t) = \textstyle\sum_{k,l} \mu_{2k} \pi(2, \boldsymbol{k}; t) + \sum_{k,l} (\mu_{2k} - \lambda_{2k})\, u(2, \boldsymbol{k}; t) \tag{D.7}$$

$$p(2;t) = p_{20}[1 + \pi(2;t)] = p_{20}\big[1 + \textstyle\sum_{k,l} \mu_{2k} \pi(2, \boldsymbol{k}; t) + (\mu_{2k} - \lambda_{2k}) u(2, \boldsymbol{k}; t)\big] \tag{D.8}$$

(D.7) describes fluctuations $\pi(2;t)$ (D.4) of mean square price $p(2;t)$ as function of partial square price disturbances $\pi(2,\boldsymbol{k};t)$ (D.6) and disturbances of squares volume $u(2,\boldsymbol{k};t)$ (D.1).

Now let's study simple model that can describe volume, value and price disturbances (D.1, D.4, D.8). Assume that $U_k^2$, $C_k^2$, $Et_{Uk}$ and $Et_{Ck}$ in (D.1, D.2) are constant or their changes are slow to compare with changes of the small dimensionless disturbances $u(2,\boldsymbol{k};t)$, $c(2,\boldsymbol{k};t)$, $et_q(\boldsymbol{k};t)$ and $et_c(\boldsymbol{k};t)$. We assume that in the linear approximation by perturbations $F_U$, $F_C$ in (5.1) and $W_U$, $W_C$ in (5.5) take form:

$$F_U(\boldsymbol{k};t) = a_{uk} Et_{U\boldsymbol{k}} \big(1 + et_u(\boldsymbol{k};t)\big) \ ; \ F_C(\boldsymbol{k};t) = a_{ck} Et_{C\boldsymbol{k}} \big(1 + et_c(\boldsymbol{k};t)\big) \tag{D.9}$$

$$W_U(\boldsymbol{k};t) = b_{uk} U_{\boldsymbol{k}}^2 \big(1 + u(2, \boldsymbol{k}; t)\big) \ \ ; \ W_C(\boldsymbol{k};t) = b_{ck} C_{\boldsymbol{k}}^2 \big(1 + c(2, \boldsymbol{k}; t)\big) \tag{D.10}$$

Taking into account (B.20, B.21) and (D.1, D.2) equations (5.1, 5.6) in linear approximation by disturbances take form:

$$U_{\boldsymbol{k}}^2 \frac{d}{dt} u(2, \boldsymbol{k}; t) = a_{uk} Et_{U\boldsymbol{k}} et_u(\boldsymbol{k};t) \ \ ; \ C_{\boldsymbol{k}}^2 \frac{d}{dt} c(2, \boldsymbol{k}; t) = a_{ck} Et_{C\boldsymbol{k}} et_c(\boldsymbol{k};t) \tag{D.11}$$

$$Et_{U\boldsymbol{k}} \frac{d}{dt} et_U(\boldsymbol{k};t) = b_{uk} U_{\boldsymbol{k}}^2 u(2, \boldsymbol{k}; t) \ ; \ Et_{C\boldsymbol{k}} \frac{d}{dt} et_C(\boldsymbol{k};t) = b_{ck} C_{\boldsymbol{k}}^2 c(2, \boldsymbol{k}; t) \tag{D.12}$$

For

$$\omega_{uk}^2 = -a_{uk}\, b_{uk} > 0 \ \ ; \quad \omega_{ck}^2 = -a_{ck}\, b_{ck} > 0 \tag{D.13}$$

simple solutions for disturbances $u(2,\boldsymbol{k};t)$, $c(2,\boldsymbol{k};t)$, $et_u(\boldsymbol{k};t)$, $et_c(\boldsymbol{k};t)$ take harmonic oscillations:

$$u(2,\boldsymbol{k};t) = u_{1k} sin\omega_{uk}t + u_{2k} cos\omega_{uk}t \quad ; \quad u_{1k}, u_{2k} \ll 1 \tag{D.14}$$

$$c(2,\boldsymbol{k};t) = c_{1k} sin\omega_{ck}t + c_{2k} cos\omega_{ck}t \quad ; \quad c_{1k}, c_{2k} \ll 1 \tag{D.15}$$

Solutions (D.8, D.9) are similar to relations derived for the first-degree trades disturbances (Olkhov, 2019b). For the case

$$\gamma_{uk}^2 = -a_{uk}\, b_{uk} < 0 \quad ; \quad \gamma_{ck}^2 = -a_{ck}\, b_{ck} < 0 \tag{D.16}$$

equations (D.11, D.12) describe exponential growth or dissipations of perturbations

$$u(2,\boldsymbol{k};t) \sim \exp \gamma_{uk}t \;\; ; c(2,\boldsymbol{k};t) \sim \exp \gamma_{ck}t \tag{D.17}$$

and perturbations $\pi(2,\boldsymbol{k};t)$ (D.6) also may follow exponential growth or decline. For different expectations $\boldsymbol{k}=(k,l)$ (4.1, B.7, B.8) solutions for disturbances $u(2,\boldsymbol{k};t)$, $c(2,\boldsymbol{k};t)$ can take (D.14, D.15) or (D.17). Combinations of perturbations determined by (D.14, D.15) or (D.17) define perturbations of price $p(2;t)$ (D.4). Thus simple equations (5.1, 5.5) on the second-degree volume and value may describe unexpected exponential growth of price $p(2;t)$ perturbations $\pi(2;t)$ and hence exponential growth of price volatility $\sigma_p^2$ (2.10).

Now let's assume that $u(2,\boldsymbol{k};t)$ and $c(2,\boldsymbol{k};t)$ disturbances depend on disturbances of expectations $ex_u(\boldsymbol{k};t)$, $ex_c(\boldsymbol{k};t)$ (see B.20, B.21):

$$Ex_U(\boldsymbol{k};t) = Ex_{U\boldsymbol{k}}\big(1 + ex_u(\boldsymbol{k};t)\big); \; Ex_C(\boldsymbol{k};t) = Ex_{C\boldsymbol{k}}\big(1 + ex_c(\boldsymbol{k};t)\big) \tag{D.18}$$

Let's take into account (D.1) and in the linear approximation by disturbances obtain:

$$Et_U(\boldsymbol{k};t) = Ex_{U\boldsymbol{k}} U_{\boldsymbol{k}}^2\big(1 + ex_u(\boldsymbol{k};t) + u(2,\boldsymbol{k};t)\big) \tag{D.19}$$

$$Et_C(\boldsymbol{k};t) = Ex_{C\boldsymbol{k}} U_{\boldsymbol{k}}^2\big(1 + ex_c(\boldsymbol{k};t) + c(2,\boldsymbol{k};t)\big) \tag{D.20}$$

Now use (D.19, D.20) and replace (D.9) by (D.21, D.22):

$$F_U(\boldsymbol{k};t) = a_{uk} Et_{U\boldsymbol{k}} U_{\boldsymbol{k}}^2\big(1 + ex_u(\boldsymbol{k};t) + u(2,\boldsymbol{k};t)\big) \tag{D.21}$$

$$F_C(\boldsymbol{k};t) = a_{ck} Et_{C\boldsymbol{k}} C_{\boldsymbol{k}}^2\big(1 + ex_c(\boldsymbol{k};t) + c(2,\boldsymbol{k};t)\big) \tag{D.22}$$

It is easy to show that for

$$\omega_{uk}^2 = -\frac{a_{uk}\, b_{uk}}{U_{\boldsymbol{k}}} > 0 \quad ; \quad \omega_{ck}^2 = -\frac{a_{ck}\, b_{ck}}{C_{\boldsymbol{k}}} > 0 \tag{D.23}$$

$$\gamma_{uk} = \; a_{uk}\frac{Ex_{Uk}}{U_{\boldsymbol{k}}} \; ; \; \gamma_{ck} = \; a_{ck}\frac{Ex_{Ck}}{C_{\boldsymbol{k}}} \tag{D.24}$$

equations (5.1, 5.5, D.10, D.21, D.22) describe damped harmonic oscillators:

$$\left[\frac{d^2}{dt^2} + \gamma_{uk}\frac{d}{dt} + \omega_{uk}^2\right]u(\boldsymbol{k};t) = 0 \;\; ; \;\; \left[\frac{d^2}{dt^2} + \gamma_{ck}\frac{d}{dt} + \omega_{ck}^2\right]c(\boldsymbol{k};t) = 0$$

$$u(\boldsymbol{k};t) \sim \sin\varphi_{uk}t \exp -\frac{\gamma_{uk}}{2}t \;\; ; \;\; c(\boldsymbol{k};t) \sim \sin\varphi_{ck}t \exp -\frac{\gamma_{ck}}{2}t \tag{D.25}$$

$$\varphi_{uk}^2 = \omega_{uk}^2 - \frac{\gamma_{uk}^2}{4} \;\; ; \;\; \varphi_{ck}^2 = \omega_{ck}^2 - \frac{\gamma_{ck}^2}{4}$$

If $\gamma_{uk}$ >0 and $\gamma_{ck}$ >0 than (D.25) and (D.4) describe trade and price *p(2;t)* disturbances as harmonic oscillations with dissipating amplitudes. But if $\gamma_{uk}$ <0 or $\gamma_{ck}$ <0 than (D.25) and *p(2;t)* in (D.4) and volatility (5.6) have harmonic modes with exponentially growing amplitudes. It is interesting that perturbations of expectations $Ex_u(\boldsymbol{k};t)$, $Ex_c(\boldsymbol{k};t)$ (D.18) can generate disturbances of trades (D.25), price *p(2;t)* (D.4) and volatility $\sigma_p^2$ (5.6) like harmonic oscillations with amplitude growth by exponent in time. It is obvious that such exponential rise of perturbations $\pi(2;t)$ and price volatility $\sigma_p^2$ (2.8) is reduced by applicability of model equations (5.1, 5.5) and relations (D.1, D.2, D.4, D.14, D.15, D.17). This approximation has sense till perturbations remain small to compare with 1.

Relations (D.4, D.8) describe mean square price *p(2;t)* (3.19) disturbances. We derive similar relations on mean price *p(1;t)* disturbances (Olkhov, 2019b):

$$p(1;t) = p_{10}\left[1 + \sum_{k,l=1}^{K}(\mu_{1\boldsymbol{k}}c(1,\boldsymbol{k};t) - \lambda_{1k}u(1,\boldsymbol{k};t))\right] \quad \text{(D.26)}$$

Coefficients $p_{10}$, $\mu_{1k}$, $\lambda_{1k}$ and functions $c(1,\boldsymbol{k};t)$, $u(1,\boldsymbol{k};t)$ (D.26) describe price *p(1;t)* and the first-degree value and volume in a way similar to $p_{10}$, $\mu_{1k}$, $\lambda_{1k}$ (D.4) and functions $c(1,\boldsymbol{k};t)$, $u(1,\boldsymbol{k};t)$ (D.5) for *p(2;t)* (D.4) and the second-degree value and volume (D.1). Disturbances of *p(2;t)* (D.4) and *p(1;t)* (D.26) describe price volatility $\sigma_p^2$ (2.10) in linear approximation by disturbances as:

$$\sigma_p^2(t) = \sigma_{p0}^2\{1 + \sum_{k,l=1}^{K} \sigma_2[\mu_{2\boldsymbol{k}}c(2,\boldsymbol{k};t) - \lambda_{2k}u(2,\boldsymbol{k};t)] - \sigma_1[\mu_{1\boldsymbol{k}}c(1,\boldsymbol{k};t) - \lambda_{1k}u(1,\boldsymbol{k};t)]\}$$

$$\sigma_{p0}^2 = p_{20} - p_{10}^2 \ ; \quad \sigma_2 = \frac{p_{20}}{\sigma_{p0}^2} \ ; \quad \sigma_1 = 2\frac{p_{10}^2}{\sigma_{p0}^2} \quad \text{(D.27)}$$

In (D.27) constants and functions with index 2 describe price *p(2;t)* (2.5; 2.6) for $n=2$ and its disturbances and index 1 describes price *p(1;t)* (1.3) (Olkhov, 2019b) and its disturbances.